%% file: Main.tex
\documentclass[acmsmall,screen]{acmart}

\def\BibTeX{{\rm B\kern-.05em{\sc i\kern-.025em b}\kern-.08emT\kern-.1667em\lower.7ex\hbox{E}\kern-.125emX}}

\input{preamble}

\begin{document}

\title{Demystifying the Silence of Correctness Bugs in PyTorch Compiler}

\author{Meiziniu Li}
\orcid{0000-0001-5947-4030}  
\affiliation{
  \institution{The Hong Kong University of Science and Technology}
  \country{China}, and 
  \institution{Guangzhou HKUST Fok Ying Tung Research Institute}
  \country{China}
}
\email{mlick@cse.ust.hk}

\author{Dongze Li}
\orcid{0009-0004-6482-8509}  
\affiliation{
  \institution{The Hong Kong University of Science and Technology}
  \country{China}, and 
  \institution{Guangzhou HKUST Fok Ying Tung Research Institute}
  \country{China}
}
\email{dongze.li@connect.ust.hk}

\author{Jianmeng Liu}
\orcid{0009-0004-8423-7781}  
\affiliation{
  \institution{Carnegie Mellon University}
  \country{USA}
}
\email{jianmenl@cs.cmu.edu}

\author{Shing-Chi Cheung*}
\orcid{0000-0002-3508-7172}  
\thanks{* Corresponding author.}
\affiliation{
  \institution{The Hong Kong University of Science and Technology}
  \country{China}, and 
  \institution{Guangzhou HKUST Fok Ying Tung Research Institute}
  \country{China}
}
\email{scc@cse.ust.hk}

\begin{abstract}
Performance optimization of AI infrastructure is key to the fast adoption of large language models (LLMs).
The PyTorch compiler (\subject{}), a core optimization tool for deep learning (DL) models (including LLMs), has received due attention.
However, \subject{} is prone to correctness bugs, which cause incorrect outputs of compiled DL models without triggering exceptions, crashes, or warnings.
These bugs pose a serious threat to the reliability of downstream LLM applications.
Data from the PyTorch community shows that 19.2\% of high-priority issues are incorrect outputs of compiled DL models induced by \subject{} bugs, the second-most-common bug category (only behind program crashes at 19.57\%).
However, no systematic study has been conducted to specifically characterize and thereby detect these bugs.
In this paper, we present the first empirical study of the correctness bugs in \subject{}, examine their characteristics, and assess the effectiveness of existing fuzzers in detecting them.
Based on our findings, we propose a proof-of-concept testing technique named \toolname{}, tailored specifically for detecting correctness bugs in \subject{}.
\toolname{} incorporates bug characteristics distilled from our empirical study, applying LLM-based test mutation to existing test cases for correctness bug detection.
At the time of writing, \toolname{} has successfully detected 23 new correctness bugs in recent \subject{}.
All these bugs have been confirmed or fixed by the PyTorch development team, and over half (14/23) of them are even marked as high-priority bugs, underscoring the usefulness of our technique.
\end{abstract}

\maketitle

\input{1-Introduction}

\input{2-Background}
\input{3-Study}
\input{5-Benchmark}
\input{6-Evaluation}

\input{7-Discussions}

\input{8-Related-Works}
\input{9-Conclusion}

\bibliographystyle{ACM-Reference-Format}
\bibliography{reference}
\end{document}

%% file: preamble.tex
\usepackage{amsmath}
\usepackage{amsfonts}
\usepackage{algorithmic}
\usepackage{graphicx}
\usepackage{textcomp}
\usepackage{xspace}
\usepackage{caption}
\usepackage{subcaption}
\usepackage{hyperref}
\usepackage{listings}
\usepackage[table]{xcolor}
\usepackage[utf8]{inputenc}
\usepackage{lipsum} 
\usepackage{xifthen}
\usepackage{pifont} 
\usepackage{array}  
\usepackage{booktabs}
\usepackage{multirow}
\usepackage{enumitem}
\usepackage{listing}
\usepackage{float}
\usepackage{ragged2e}
\usepackage{tabularray}
\usepackage{cleveref}
\UseTblrLibrary{booktabs}

\newfloat{listing}{htbp}{lst}
\floatname{listing}{Listing}

\usepackage{tcolorbox}
\tcbuselibrary{skins}

\newcommand*{\toolname}{\textsc{AlignGuard}\xspace}
\newcommand*{\subject}{\texttt{torch.compile}\xspace}

\definecolor{codegreen}{rgb}{0,0.6,0}
\definecolor{codegray}{rgb}{0.5,0.5,0.5}
\definecolor{codepurple}{rgb}{0.58,0,0.82}
\definecolor{backcolour}{rgb}{0.95,0.95,0.92}

\lstdefinestyle{mystyle}{
    backgroundcolor=\color{backcolour},   
    basicstyle=\scriptsize\ttfamily,
    breaklines=true,
    frame=single,
    framesep=0.5ex,
    xleftmargin=0.5ex,
    escapeinside=||
}

\definecolor{codegreen}{rgb}{0,0.6,0}
\definecolor{codegray}{rgb}{0.5,0.5,0.5}
\definecolor{codepurple}{rgb}{0.58,0,0.82}
\definecolor{backcolour}{rgb}{0.98,0.98,0.98}
\definecolor{bg}{rgb}{0.95,0.95,0.95}

\tcbset{
    MyFrame/.style={
        before={\noindent},
        after={\par},
        left=5pt,
        right=5pt,
        colframe=black,
        boxrule=0.2pt,
        arc=5pt,
        top=5.5pt,
        bottom=5.5pt,
        colback=bg,
        before skip=5.5pt,
        after skip=5.5pt,
    }
}

%% file: 1-Introduction.tex
\section{Introduction}\label{sec:introduction}
The exponential growth in deep learning (DL) model scale has made optimization essential for efficient model training and inference, especially for large language models (LLMs) that require extensive computational resources.
PyTorch compiler, named \subject{}, is receiving growing attention as a core optimization tool for LLMs, with its ability to enable graph optimization without sacrificing the dynamic programming flexibility of Python~\cite{torchcompile_techreport}.
Many AI infrastructures for LLMs and leading organizations, such as vLLM~\cite{vllm} (with over 74k GitHub stars) and AWS~\cite{aws_torchcompile}, have integrated \subject{} to elevate the efficiency of LLMs.

\begin{figure}[htbp]
    \centering
    \includegraphics[width=0.5\linewidth]{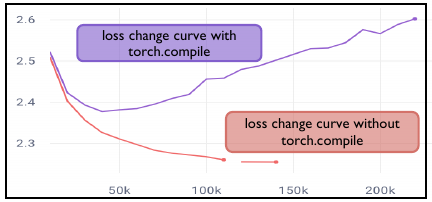}
    \caption{An industrial LLM fails to converge during training due to a correctness bug in \subject{}.}\label{fig:sec2_silenterror_example}
    \Description{A correctness bug in \subject{} causes an industrial LLM to fail to saturate during training.}
\end{figure}

However, \subject{} suffers from correctness bugs, which are bugs that lead to incorrect outputs of the compiled DL models, without raising exceptions, crashes, or warnings.
These bugs compromise the reliability of mission-critical AI applications built upon compiled DL models, such as medical diagnosis~\cite{llm4disease_diagnosis}, financial risk control~\cite{llm4financial_control}, and autonomous driving~\cite{yang2023llm4drive,fu2024drive}, where incorrect model outputs could have catastrophic consequences. 
Figure~\ref{fig:sec2_silenterror_example} illustrates a severe consequence of such bugs: an industrial LLM (OLMo~\cite{olmo20242olmo2furious}) fails to converge during training due to a correctness bug in \subject{}, resulting in a drastic performance drop~\cite{olmo_bug}.
Indeed, our analysis of PyTorch's GitHub issues highlights that 19.2\% of high-priority issues in the past year are related to correctness bugs in \subject{}—the second most dominant bug type, only second to program crashes (19.57\%) (see Section~\ref{sec:background}).

Although prior works have been proposed to explore the bug characteristics in DL compilers~\cite{dlcompilerbugstudy1,dlcompilerbugstudy2}, they focus on general bug patterns rather than specifically targeting the correctness bugs.
Moreover, these studies focus on static-graph DL compilers, such as TVM~\cite{tvm} and Glow~\cite{glow}, whose inputs are computational graphs pre-defined by compiler users before execution.
In contrast, \subject{} supports the dynamic Python imperative code as input and performs just-in-time (JIT) compilation via dynamically capturing graphs during runtime.
This fundamental architectural difference leads to distinct input patterns and erroneous behaviors in \subject{} (see Section~\ref{subsec:sec2_existing_studies}).

This research gap motivates us to perform the first empirical study on correctness bugs in the PyTorch compiler, aiming to facilitate a deeper understanding and more effective detection of these bugs.
Firstly, we collect and manually inspect 116 real-world correctness bugs in \subject{}, and systematically characterize their key properties, including bug types, root causes, and triggering patterns.
Our analysis identifies three main bug types (i.e., \textit{graph-related bugs}, \textit{operator-related bugs}, and \textit{memory-related bugs}) with six subcategories, and further derive explicit root causes and triggering patterns for each subcategory.
We obtain several interesting findings, for instance, \textit{in-place operation handling bugs} are the most prevalent subcategory, accounting for 21.6\% in our dataset, and these bugs are commonly caused by the incorrect tracking of aliasing relationships inside \subject{}.

Leveraging our analysis results, we further conduct a systematic evaluation of the five state-of-the-art DL compiler testing techniques, assessing their ability to detect correctness bugs in \subject{}.
Overall, we find that the selected techniques collectively detect only 33.8\% of correctness bugs in our benchmark, highlighting a critical gap in detecting such bugs in \subject{}.
A key reason behind this gap is the lack of domain-specific test generation strategies tailored to the unique triggering patterns of correctness bugs in \subject{}.
For instance, \textit{graph-related bugs} are often triggered by non-computational operations in DL models, yet such operations are not covered by existing techniques.
As a result, none of these \textit{graph-related bugs} can be successfully detected by the selected techniques.

To improve the detection of correctness bugs in \subject{}, we present \toolname{}, a proof-of-concept LLM-based test mutation approach designed specifically for exposing correctness bugs in \subject{}.
Leveraging the bug characteristics distilled from our empirical study, \toolname{} performs mutations on existing test cases to induce the triggering patterns of correctness bugs in \subject{}, aiming for more effective detection of these bugs.
\toolname{} has successfully detected 23 previously unknown correctness bugs, all of which have been confirmed and 10 reported bugs have been fixed by PyTorch developers.
Notably, \textbf{more than half (14/23) of the reported bugs have been labeled as high-priority issues}, demonstrating the usefulness of \toolname{}.

To sum up, we make three major contributions.

\begin{itemize}
    \item We present the first empirical study on correctness bugs in PyTorch compiler based on 116 collected real-world cases.
    We systematically characterize the bug types, root causes, and triggering patterns for these bugs.
    \item We conduct the first experiment to evaluate the effectiveness of five state-of-the-art DL compiler testing techniques in detecting PyTorch compiler correctness bugs.
    By analyzing the characteristics of detected and missed correctness bugs, we identify the strengths and weaknesses of these techniques in this task and suggest directions for future improvement.
    \item We develop a testing technique named \toolname{}, which incorporates the bug characteristics distilled from our study to enhance the detection of correctness bugs in \subject{}.
    \toolname{} has successfully detected 23 new correctness bugs in \subject{}.
    All 23 bugs have been confirmed or fixed by PyTorch developers, including \textbf{14 marked as high-priority bugs.}
\end{itemize}

%% file: 2-Background.tex
\section{Background \& Motivation}\label{sec:background} 
This section introduces the background of the PyTorch compiler and explains the motivation behind our study on its correctness bugs.

\subsection{Overview of PyTorch Compiler}\label{subsec:sec2_overview_tc}
PyTorch compiler (\subject{}) is a critical infrastructure that optimizes the execution of DL models across hardware platforms.
It takes PyTorch-based DL models as input, applies multi-stage optimizations (e.g., operator fusion, memory scheduling), and generates efficient hardware-specific executables~\cite{zhang2025deeplearninglibrarytesting}.

\begin{figure}[htbp]
    \centering
    \includegraphics[width=\linewidth]{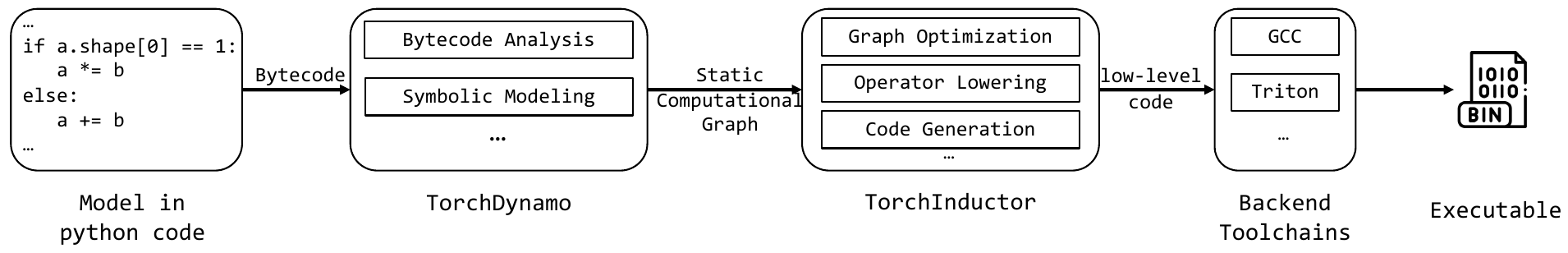}
    \caption{Architecture of PyTorch compiler.}\label{fig:sec2_dlcompiler_stack}
    \Description{A diagram showing the typical deep learning compiler stack.}
\end{figure}

As illustrated in Figure~\ref{fig:sec2_dlcompiler_stack}, \subject{} includes three core components:
\begin{itemize}[leftmargin=*]
    \item \textbf{TorchDynamo}: 
    TorchDynamo serves as the frontend of \subject{}, responsible for capturing computational graphs (FX graphs) from user-defined PyTorch models.
    Specifically, it dynamically analyzes Python bytecode to extract PyTorch operations into an FX graph, which is then passed to the backend compiler (e.g., TorchInductor) for just-in-time (JIT) optimization.
    To support the flexibility of Python, TorchDynamo uses symbolic modeling and other dedicated mechanisms to handle Python-native features such as side effects, dynamic control flows, and closures.
    \item \textbf{TorchInductor}: 
    TorchInductor is the official backend compiler that lowers the FX graph into efficient low-level code.
    It performs a series of optimizations and translates optimized representations into hardware-specific low-level code, such as OpenAI Triton code for GPUs and C++ code for CPUs~\cite{torchcompile_techreport,triton}.
    \item \textbf{Backend Toolchains}: Finally, \subject{} leverages the backend toolchains to compile the generated low-level code into machine-executable binaries.
    For CPUs, it uses GCC/Clang; for GPUs, it uses the Triton compiler to produce high-performance kernels~\cite{triton}.
\end{itemize}

\subsection{The Significance of PyTorch Compiler and Its Correctness Bugs}\label{subsec:sec2_significance_tc}
\begin{figure}[htbp]
    \centering
    \begin{minipage}[b]{0.9\textwidth}
        \centering
        \includegraphics[width=0.9\textwidth]{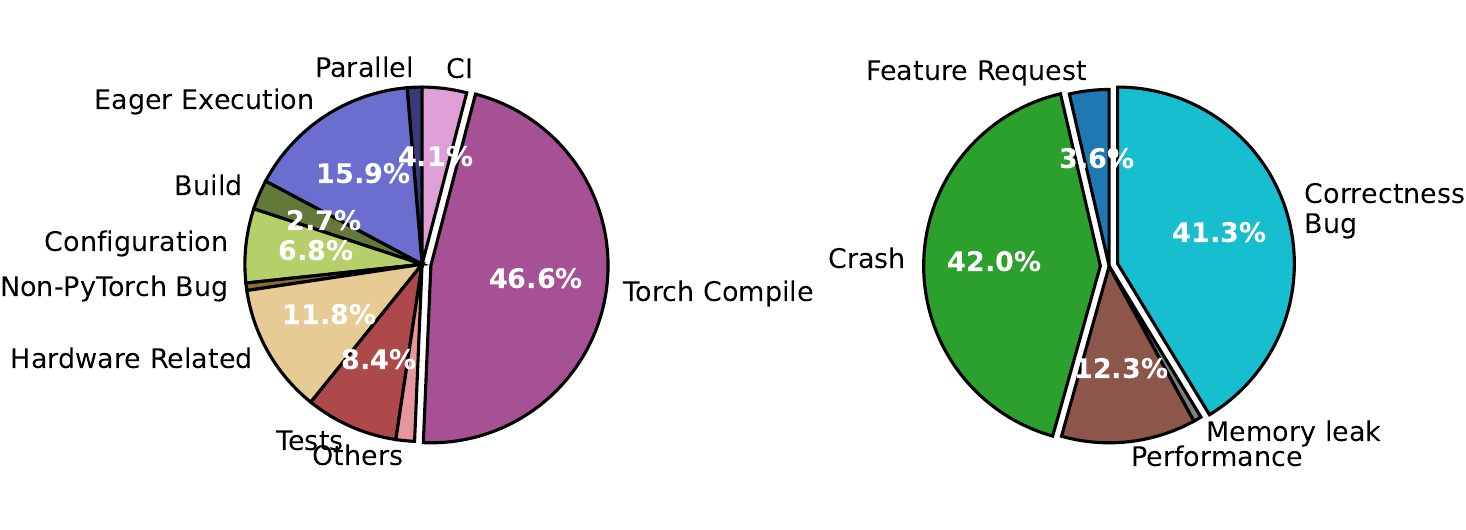}
    \end{minipage}

    \vspace{-0.2cm}

    \centering
    \begin{minipage}[b]{0.9\textwidth}
        \centering
        \includegraphics[width=0.9\textwidth]{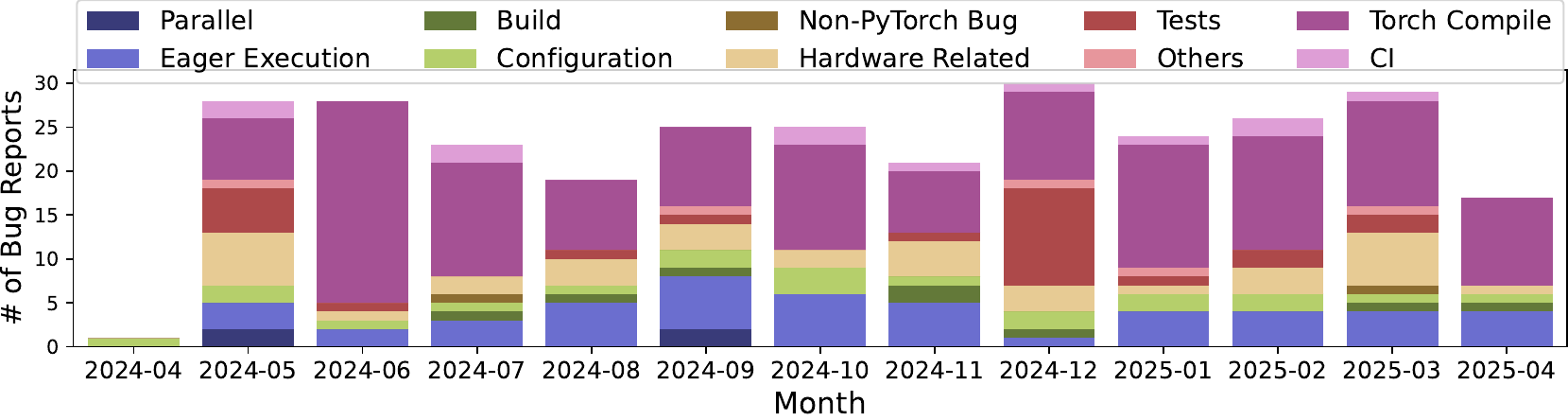}
    \end{minipage}

    \caption{Distribution of faulty components (bottom and top left) and symptoms of high-priority issues in \subject{} (top right).
    }\label{fig:sec2_highpri_location}
    \Description{Two bar charts showing the distribution of faulty components and symptoms of high-priority issues in PyTorch compiler.}
\end{figure}

As a core and widely adopted component in the PyTorch ecosystem, \subject{} and its reliability have received considerable attention from both the user community and the PyTorch development team.
To quantify the prevalence and severity of \subject{} issues, we collect and analyze recent high-priority bug reports, which are critical issues explicitly labeled by the PyTorch team for extra attention.
Among the 296 high-priority reports collected, 46.6\% (138/296) are discussing bugs in \subject{}, making it the most frequently reported faulty component in PyTorch across nearly every month between April 19, 2024 and April 19, 2025 (see Figure~\ref{fig:sec2_highpri_location}).

We further break down the failure symptoms of these high-priority \subject{} issues.
Correctness bugs (incorrect outputs of compiled models) stand out as the second most common symptom.
These bugs account for 41.3\% (57/138) of all cases, only 0.7\% lower than program crashes (see the top-right of Figure~\ref{fig:sec2_highpri_location}).
Notably, the ratio of correctness bugs is substantially higher than that reported in a prior empirical study on DL compiler bugs~\cite{dlcompilerbugstudy1}, which reported that only 26.3\% (151/575) of DL compiler bugs would induce incorrect outputs.
This discrepancy highlights that correctness bugs are a particularly critical and understudied problem in \subject{}.

Two key factors contribute to the growing attention to correctness bugs in \subject{}.
Firstly, these bugs could severely degrade the performance of LLM applications built upon the compiled DL models, wasting substantial developer effort and computational resources.
Figure~\ref{fig:sec2_silenterror_example} presents an example of severe consequences caused by correctness bugs: an industrial LLM (OLMo~\cite{olmo20242olmo2furious}) fails to converge during the training phase due to a real-world compilation bug in \subject{}, leading to significant performance degradation~\cite{olmo_bug}.
Secondly, these bugs are inherently difficult to detect and debug due to their silent nature (i.e., no program crash or error message)~\cite{jiang2025trainingconfidencecatchingsilent}, especially when the deployed DL models (notably LLMs) are large-scale and complex.
Accordingly, identifying and exposing correctness bugs in \subject{} in a timely manner is critical to ensuring the reliability of AI applications built on DL models compiled by \subject{}.

\subsection{Research Gaps}\label{subsec:sec2_existing_studies}
Although existing research~\cite{dlcompilerbugstudy1, dlcompilerbugstudy2} has extensively investigated bugs inside DL compilers (e.g., TVM~\cite{tvm}, Glow~\cite{glow}, and nGraph~\cite{ngraph}), a substantial gap remains in understanding the unique characteristics of correctness bugs in \subject{}.
We identify two key limitations in existing studies that motivate our research.

Firstly, existing studies focus on general bug symptoms and root causes of DL compilers, rather than specifically targeting correctness bugs.
While correctness bugs are also included in the scope of these studies, they are not the focus of dedicated analysis, leaving the underlying root causes and bug-triggering patterns of these bugs underexplored.
Secondly, all previously studied compilers (e.g., TVM) target static computational graphs that are fully defined before model execution~\cite{torchcompile_techreport}.
In contrast, \subject{} takes dynamic Python imperative code as input and performs just-in-time (JIT) compilation by dynamically capturing graphs during runtime~\cite{torchcompile_techreport}.
This fundamental architectural difference leads to distinct input patterns and erroneous behaviors in \subject{}.
For example, correctness bugs introduced during dynamic graph capture and Python bytecode transformation have not been systematically studied.
These limitations motivate a systematic study on correctness bugs in \subject{}.
To bridge this research gap, we conduct the first empirical study on correctness bugs in \subject{}, aiming to reveal the characteristics of these bugs, assess the performance of state-of-the-art fuzzers, and provide empirical evidence for designing effective fuzzing techniques for detecting these bugs in \subject{}.

%% file: 3-Study.tex
\section{Empirical Study}\label{sec:empirical_study}

Following the common practice in existing empirical studies on software bugs~\cite{taxonomy_real_faults,ma2025comprehensive,YANG2022107004}, we design our empirical study by first categorizing the correctness bugs in \subject{} into different bug types, based on their faulty components, and then performing a fine-grained analysis of each bug type.
Such a design is motivated by our observation that \subject{} is a complex system comprising multiple components, where bugs in different components may exhibit distinct root causes and bug-triggering patterns.
Therefore, conducting a fine-grained analysis of each bug type enables us to identify the unique characteristics of correctness bugs in \subject{} and provide more concrete actionable insights for their detection.

The remainder of this section is organized as follows. 
Section~\ref{subsec:error_collection_methodology} details our methodology for collecting correctness bugs in \subject{}, including automated data collection and manual inspection. 
Section~\ref{subsec:error_types} presents the results of bug categorization and analysis, which reveal the distribution of bug types in \subject{}, along with their respective root causes and bug-triggering patterns.

\subsection{Bug Collection}\label{subsec:error_collection_methodology}
\textbf{Data Collection.}
To collect correctness bugs in \subject{}, we systematically collect closed issues related to \subject{} from the issue tracker of PyTorch's GitHub repository~\footnote{\url{https://github.com/pytorch/pytorch/issues}}.
Only closed issues are selected because they include bugs that have been confirmed, investigated, and fixed by PyTorch developers, thus providing comprehensive contextual information (e.g., developer discussions, code snippets, version changes, and links to related fixes) to facilitate bug analysis.
We performed a keyword-based search to extract candidate issues containing the term `torch.compile', covering the period from April 19, 2023, to April 19, 2025 (our data collection cutoff date).
Using `torch.compile' as the keyword enables us to focus on issues directly relevant to \subject{} and capture most \subject{}-related issues, as \subject{}-related issues typically include `torch.compile' (i.e., the API name for compiling DL models) in their reproducible code and bug descriptions.
This search yielded a total of 2,542 candidate issues, which served as the initial pool for our subsequent filtering process.

\input{tables/Study-Data-Collection.tex}

\textbf{Manual Filtering.}
To identify real correctness bugs from the candidate pool, we conduct a multi-stage manual filtering process (detailed in Table~\ref{tab:sec3_data_collection}), guided by a set of filtering criteria adapted from a prior empirical study~\cite{silenterrorinDLframework} on DL framework silent bugs.
Each criterion excludes a specific category of irrelevant issues, and the detailed filtering steps for each criterion are elaborated below.

\begin{itemize}[leftmargin=*]
    \item \textbf{Non-bug issues (step 1)}:
    We exclude non-bug issues, including feature requests and general help requests.
    Specifically, an issue that lacks confirmation from PyTorch developers (e.g., via comments, linked pull requests, or bug-specific labels) is considered a non-bug issue.
    \item \textbf{Non-\subject{} issues (step 2)}: 
    We exclude issues unrelated to the \subject{} component. 
    The relevance is determined based on evidence such as developer discussions, linked pull requests that modify \subject{} code, or the \subject{}-specific issue label (i.e., `oncall: pt2').
    \item \textbf{Not correctness bugs (step 3)}:
    We exclude issues that do not exhibit the symptom of correctness bugs.
    Specifically, we carefully identify their symptoms based on each issue's description and developer discussion, including only those that report incorrect outputs of compiled models.
    This step excludes issues that report other types of bugs, such as exceptions, program crashes, or memory leaks.
    \item \textbf{Not confirmed by developers (step 4)}:
    We filter out issues that have not been confirmed by PyTorch developers.
    This includes filtering issues that are explicitly commented by developers as expected behavior, or closed by developers without fixing them.
    \item \textbf{Duplicated issues (step 5)}:
    We exclude duplicate issues that are either explicitly marked as duplicates by PyTorch developers or linked to a previously reported original issue without an additional fix.
    \item \textbf{Unclear root causes (step 6)}:
    We exclude issues that remain unresolved, lack a clear description of their root cause, or are not linked to fixing commits in the issue discussion.
\end{itemize}

As a result, we identify 116 real correctness bugs in \subject{} for our subsequent analysis.

\subsection{{Bug Types}}\label{subsec:error_types}
\begin{figure}[htbp]
    \centering
    \includegraphics[width=0.7\linewidth]{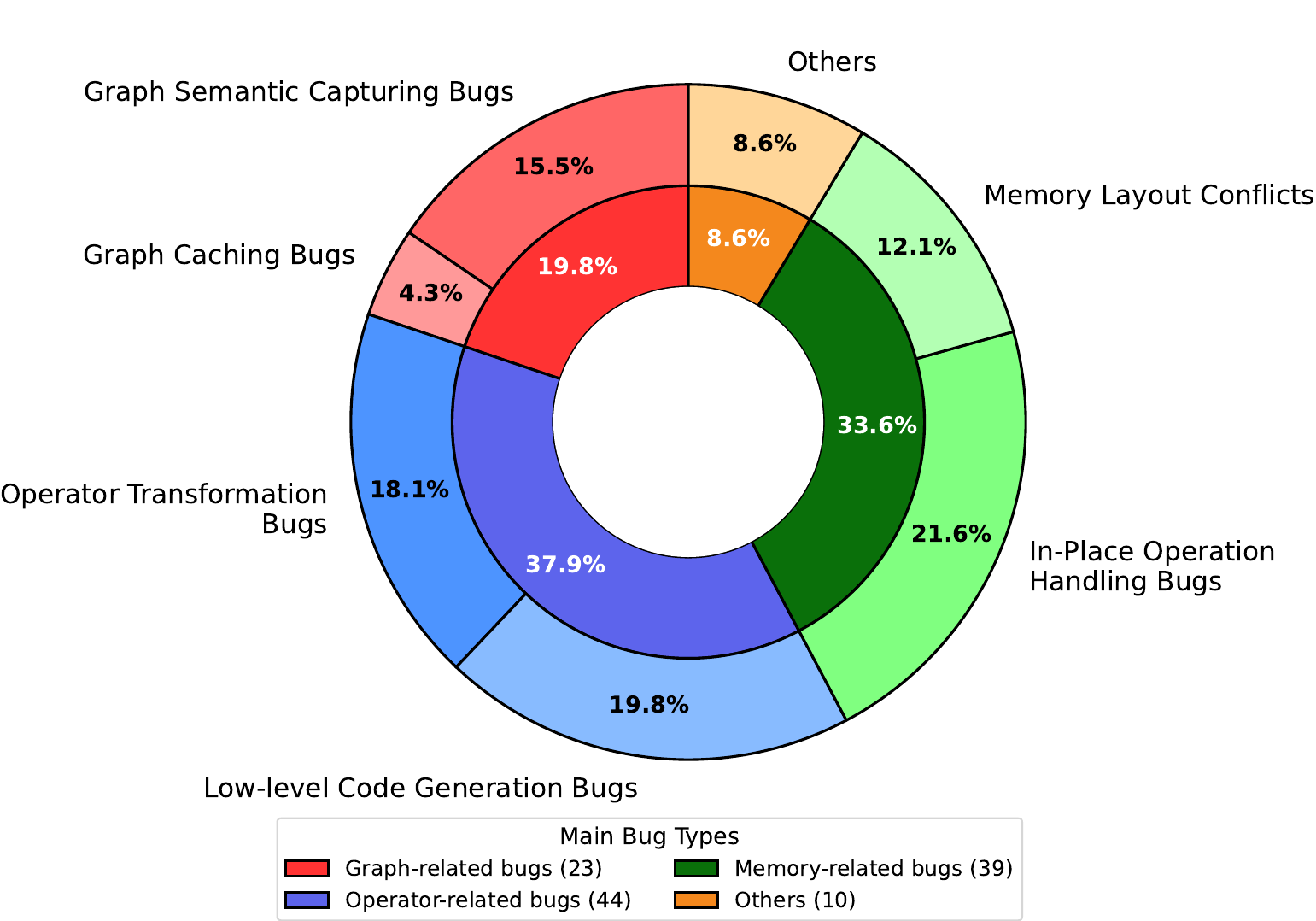}
    \caption{Distribution of Bug Types}\label{fig:sec3_error_types}
    \Description{A diagram showing the distribution of bug types in \subject{}.}
\end{figure}

We identify three main bug types of correctness bugs in \subject{}, based on the components that they infect.
Figure~\ref{fig:sec3_error_types} illustrates the distribution of these bug types, along with their respective subcategories.
Among these, \textit{operator-related bugs} (bugs in operator optimization) are the most prevalent, accounting for 37.9\% of all cases, followed by \textit{memory-related bugs} (33.6\%) (bugs in memory management) and \textit{graph-related bugs} (19.8\%) (bugs in graph construction).
This section investigates the root causes and bug-triggering patterns of each bug type and its subcategories.
Table~\ref{tab:error_characteristics_overview} offers an overview of all bug types, including their definitions, dominant root causes, and primary bug-triggering patterns.

\input{tables/Study-Error-Characteristics.tex}

\input{3-2-1-GraphRelatedError.tex}

\input{3-2-2-OperatorRelatedError.tex}

\input{3-2-3-MemoryRelatedError.tex}

\subsubsection{Other Bugs.}
We also identify several other types of correctness bugs that do not fit into the above categories, including \textit{precision bugs}, \textit{configuration bugs}, and \textit{external library bugs}.

\textbf{Precision Bugs (3/116).}
This subcategory refers to correctness bugs caused by precision issues (e.g., incorrect precision data type casting) in numerical computations within \subject{}.
Numerical precision is a crucial aspect of DL compilers, as it can significantly impact the accuracy of model outputs, especially in operations involving floating-point arithmetic.

\textbf{Configuration Bugs (5/116).}
This subcategory refers to correctness bugs caused by incorrect or suboptimal configuration settings for backend toolchains (e.g., GCC compiler flags).
For example, \subject{} uses the `-funsafe-math-optimizations' flag for the GCC compiler to optimize generated C++ code, which is inconsistent with the eager (non-compiled) execution, thus leading to different rounding errors.

\textbf{External Library Bugs (2/116).}
Two correctness bugs are caused by issues in external libraries that serve as dependencies of \subject{}.
These two bugs are both related to bugs inside the Triton compiler.
To avoid this correctness bug from the \subject{} side, PyTorch developers either apply a temporary workaround or use the patched version of the Triton compiler.

%% file: tables/Study-Data-Collection.tex
\begin{table}[htbp]
\centering
\caption{Step-by-step filtering process for identifying correctness bug issues in \subject{}}\label{tab:sec3_data_collection}
\renewcommand{\arraystretch}{1.1}
\resizebox{\linewidth}{!}{
\begin{tabular}{l l r r}
\toprule
\textbf{Step} & \textbf{Filtering Criterion / Category} & \textbf{\# Excluded} & \textbf{\# Remaining} \\
\midrule
--- & Initial issues gathered with keyword `torch.compile' & --- & 2,542 \\
\midrule
1 & Excluded Issues: Feature requests and general help requests & 581 & 1,961 \\
2 & Excluded Issues: Issues not discussing a bug within \subject{} & 379 & 1,582 \\
3 & Excluded Issues: Issues not related to correctness bugs & 1,329 & 253 \\
4 & Excluded Issues: Issues not confirmed by developers & 93 & 160 \\
5 & Excluded Issues: Duplicated issues (confirmed by developers) & 34 & 126 \\
6 & Excluded Issues: Issues fixed without clear root causes & 10 & 116 \\
\midrule
\rowcolor{gray!20}
& \textbf{Studied Dataset (Confirmed Correctness Bugs)} & --- & \textbf{116} \\
\bottomrule
\end{tabular}
}
\end{table}

%% file: tables/Study-Error-Characteristics.tex
\DefTblrTemplate{firsthead, middlehead,lasthead}{default}{}
\begin{table}[htbp]
\centering
\caption{Characteristics of correctness bugs in \subject{}.}\label{tab:error_characteristics_overview}
\resizebox{\linewidth}{!}{
  \begin{talltblr}[
    note{\ding{73}} = {High-frequency bug subcategory},
  ]{
    colspec={>{\RaggedRight}p{1.5cm} | >{\RaggedRight}p{2.5cm} | >{\RaggedRight}p{4cm} | >{\RaggedRight}p{3.5cm} | >{\RaggedRight}p{3.5cm}},
    hline{none}, vline{none},
    row{1}={font=\bfseries},
    row{2-3}={red!10}, 
    row{4-5}={blue!10},
    row{6-7}={green!10},
    rowsep=4pt,
  }
    \toprule
    Category & Subcategory & Description & Dominant Root Cause & Primary Bug-Triggering Pattern \\
    \midrule
    \SetCell[r=2]{} Graph-related Bugs (\S\ref{subsec:sec3_graph_related_error})  %
        & Graph Semantic Capturing Bugs (\ding{73})
        & Incorrect capture of front-end computational semantics       
        & Incomplete symbolic tracing for custom logic    
        & Models involving non-computational operations     \\
      \cmidrule{2-5}
        & Graph Caching Bugs              
        & Incorrect reuse of cached computational graphs    
        & Incorrect guard conditions for graph caching
        & Repeated compilation with diverse internal model states           \\
    \midrule
    \SetCell[r=2]{} Operator-related Bugs (\S\ref{subsec:sec3_operator_related_error})
        & Operator Transformation Bugs (\ding{73})
        & Incorrect transformation from high-level operators to low-level operators
        & Lack of boundary scenario handling logic    
        & Single operator with non-default parameters or edge-case inputs     \\
      \cmidrule{2-5}
        & \textbf{Low-Level Code Generation Bugs} (\ding{73}\ding{73})
        & Incorrect generation of low-level code (e.g., \texttt{Triton})
        & Inaccurate preservation of numerical properties
        & Extreme values for numerical sensitive operators           \\
    \midrule
    \SetCell[r=2]{} Memory-related Bugs (\S\ref{subsec:sec3_memory_related_error})
        & \textbf{In-Place Operation Handling Bugs} (\ding{73}\ding{73})
        & Improper handling of in-place operations
        & Incorrect tracking of aliasing relationships
        & In-place operations applied to tensor aliases or views     \\
      \cmidrule{2-5}
        & Memory Layout Conflicts             
        & Incorrect preservation of tensor memory layout    
        & Missing layout compatibility checks
        & Operator sequences inducing memory layout transformation           \\
    \bottomrule
  \end{talltblr}
}
\end{table}

%% file: 3-2-1-GraphRelatedError.tex
\subsubsection{Graph-related Bugs.}\label{subsec:sec3_graph_related_error}
Graph-related bugs account for 19.8\% (23/116) of all cases in our dataset.
These bugs occur during the front-end graph construction stage of \subject{}, where the user-defined model is transformed into a static computational graph for subsequent optimizations.
We subdivide the graph-related bugs into two subcategories:
\textit{graph semantic capturing bugs} and \textit{graph caching bugs}. 

\textbf{Bug 1: Graph Semantic Capturing Bugs (18/116).}
\textit{Graph Semantic Capturing Bugs} refer to correctness bugs caused by the incorrect capture of computational semantics when analyzing the user-defined models.
Specifically, \subject{} relies on symbolic tracing to capture the PyTorch operations from the Python bytecode of user programs, and then applies lightweight graph optimizations.
However, this tracing and optimization process may fail to reliably capture and preserve the model's intended semantics, thus resulting in incorrect static computational graphs.

\textit{\textbf{Root Causes}}.
The root causes of \textit{Graph Semantic Capturing Bugs} can be summarized into three subcategories:

\begin{itemize}[leftmargin=*]
    \item \textit{Incomplete Symbolic Tracing for Custom Logic (9/18)}:
    The symbolic tracing process fails to apply necessary tracing on user-defined operations, especially for self-defined functions not natively supported by \subject{}.
    This includes missing tracing logics (e.g., missing layout metadata propagation) for custom user-defined functions such as \texttt{autograd.Function} functions or external library APIs.
    \item \textit{Aggressive Optimization (5/18)}:
    The lightweight optimization passes (e.g., dead code elimination) incorrectly remove or alter nodes/values in the static graph.
    This includes eliminating state-mutating operations (e.g., RNG state mutation via `\texttt{torch.manual\_seed}') or essential mathematical operations (e.g., \texttt{abs} in \texttt{lgamma} for M4 devices) as `dead code', or mis-specializing symbolic values into fixed constants.
    \item \textit{Lack of Fallback Handling for Unsupported APIs (4/18)}:
    \subject{} fails to implement robust fallback mechanisms for APIs that are not fully supported by subsequent optimizations.
    Instead of triggering a fallback to preserve the original semantic of user-defined models, \subject{} silently omits those unsupported APIs in the compiled graph, leading to incorrect model outputs.
\end{itemize}

\textit{\textbf{Bug-triggering Patterns}}.
We summarize three common bug-triggering patterns for this bug type:
\begin{itemize}[leftmargin=*]
    \item \textit{Non-Computational Operations (9/18)}: 
    Bugs are triggered by user-defined models containing non-computational operations.
    These operations do not form the computational graph (e.g., not ATen operations) of the model but determine the control flow and data flow when executing user-defined models, including Python native logic (e.g,. \texttt{range}), and PyTorch framework-level utility functions.
    \subject{} fails to accurately preserve the complete semantics of these non-computational operations during tracing, leading to incorrect model outputs.
    \item \textit{Execution Context Mutation (5/18)}:
    Bugs are caused when user-defined models include operations that mutate the model's execution contexts.
    These operations include global state modifications (e.g., \texttt{TorchDispatchMode}) and compilation configuration modifications (e.g., configure \subject{} to use a customized backend).
    During the compilation, \subject{} cannot properly capture or preserve these context mutations, thus generating incorrect static computational graphs.
    \item \textit{Corner Case Scenario (4/18)}:
    Bugs are induced by models involving corner-case model inputs (e.g., empty tensors), non-default operator parameters (e.g., negative padding values), or specific execution pipelines (e.g., deep copy a model's weights and perform model training).
    These corner cases would trigger \subject{}'s incorrect handling of symbolic shapes or incorrect control flow tracing.
\end{itemize}

\textit{\textbf{Example Bug}}.
Listing~\ref{lst:error1_example} demonstrates an example of \textit{Graph Semantic Capturing Bugs}~\cite{graph_construction_error}.
In this example, \subject{} fails to capture the execution context that dispatches the \texttt{add} operation to the \texttt{mul} operation via the \texttt{TorchDispatchMode} API\@.
As a result, the actual computational graph is performing multiplication instead of addition, leading to incorrect model outputs.

\input{listings/error_examples/error1.tex}

\begin{tcolorbox}[MyFrame]
\textbf{Summary of Graph Semantic Capturing Bugs}:
\textit{Graph Semantic Capturing Bugs} account for 15.5\% (18/116) of cases in our dataset. 
These bugs occur during the front-end graph construction stage of \subject{}, where static computational graphs are incorrectly generated.
These bugs are caused by three reasons: incomplete symbolic tracing for custom logic, aggressive optimization during graph construction, and the lack of fallback handling for unsupported APIs.
Regarding the bug-triggering patterns, non-computational operations, execution context mutations, and corner case scenarios are three common patterns that trigger these bugs.
\end{tcolorbox}

\textbf{Bug 2: Graph Caching Bugs (5/116).}
\textit{Graph Caching Bugs} are defined as correctness bugs caused by the misuse of cached computational graphs.
The goal of graph caching in \subject{} is to mitigate redundant compilation processes through the reuse of cached graphs.
However, reusing semantically inequivalent cached graphs can lead to incorrect model outputs.

\textit{\textbf{Root Cause}}.
The root cause of graph caching bugs is the improper implementation of the \textit{guard condition}, which verifies the semantic equivalence between the cached graphs and the user-defined models.
Incorrect guard conditions may fail to comprehensively model all factors necessary for semantic equivalence verification, thus incorrectly reusing a semantically inequivalent graph for the user-defined model.
These missed factors include, but are not limited to: model input properties (e.g., \texttt{\_max\_seqlen} for NestedTensor inputs, tensor shapes) and mutable internal states of the model.

\textit{\textbf{Bug-triggering Pattern}}.
The bug-triggering pattern of \textit{Graph Caching Bugs} includes a specific testing pipeline: repeatedly executing the compiled user-defined model with varying internal model states or input properties.

This pipeline is specific to the graph caching mechanism in \subject{}: varying internal states or input properties between model executions force the guard conditions to be re-evaluated, and possible graph-caching bugs can thereby be exposed by checking the model's outputs through repeated executions.
Notably, this pipeline is missed by existing DL compiler fuzzers~\cite{whitefox,neuri,nnsmith,dlcompilerbugstudy1}, which predominantly compile the user-defined model once to detect correctness bugs.

\textit{\textbf{Example Bug}}.
A typical graph caching bug is illustrated in Listing~\ref{lst:error2_example}~\cite{graph_caching_error}.
Here, the compiled model is executed six times, with the internal state variable \texttt{self.value} incremented by 1 in each forward pass to index the tensor \texttt{self.cache}, so a sequentially increasing output is expected.
Nevertheless, the model's outputs become incorrect after repeated executions.
This bug occurs because the guard conditions of \subject{} fail to track the dynamic updates of \texttt{self.value}, which causes the compiler to reuse an outdated static computational graph.

\input{listings/error_examples/error2.tex}

\begin{tcolorbox}[MyFrame]
\textbf{Summary of Graph Caching Bugs}:
\textit{Graph Caching Bugs} in \subject{} are correctness bugs caused by the misuse of cached static computational graphs.
The root cause of these bugs is the incorrect guard condition that fails to model mutable internal model states and input properties.
Unlike existing DL compiler fuzzers that executed the compiled model once, detecting graph caching bugs requires repeated execution.
\end{tcolorbox}

%% file: listings/error_examples/error1.tex
\begin{listing}
\begin{lstlisting}
|\textbf{\underline{User-Defined Model}}|
class RewriteAddToMul(TorchDispatchMode):
    def __torch_dispatch__(self, func, ...):
        if func is torch.ops.aten.add.Tensor:
            func = torch.ops.aten.mul.Tensor
        return func(...)
def model(x):
    return x + x  # |\textbf{\textcolor{codegray}{Should replace `add' operator with `mul' operator}}|
x = torch.tensor([3.0])
with RewriteAddToMul():
    output = torch.compile(model)(x)
|\textbf{\underline{\textcolor{red}{Buggy Output}}}|
print(output)  # |\textbf{\textcolor{red}{tensor([6.]) -- Incorrect, `add' operator is used}}|
|\textbf{\underline{\textcolor{codegreen}{Fixed Output}}}|
print(output)  # |\textbf{\textcolor{codegreen}{tensor([9.]) -- Correct, `mul' operator is used}}|
\end{lstlisting}
\caption{Example of \textit{Graph Semantic Capturing Bugs}. In this example, \subject{} fails to capture the semantic of the \texttt{RewriteAddToMul} dispatch, thus incorrectly applies the original \texttt{add} operator instead of the rewritten \texttt{mul} operator.}
\label{lst:error1_example}
\end{listing}

%% file: listings/error_examples/error2.tex
\begin{listing}
\begin{lstlisting}
|\textbf{\underline{User-Defined Model}}|
class Model(torch.nn.Module):
    def __init__(self):
        super(Model, self).__init__()
        self.value = -1
        self.cache = torch.tensor([2, 3, 4, 5, 6, 7])
    def forward(self):
        self.value += 1
        return self.cache[self.value]
model = torch.compile(Model(),...)
output = []
for _ in range(6):
    output.append(model.forward())
|\textbf{\underline{\textcolor{red}{Buggy Output}}}|
print(output)  # |[2, 3, 4, 5, \textbf{\textcolor{red}{5, 5] -- The last two elements are incorrect}}|
|\textbf{\underline{\textcolor{codegreen}{Fixed Output}}}|
print(output)  # |[2, 3, 4, 5, \textbf{\textcolor{codegreen}{6, 7]}}|
\end{lstlisting}
\caption{Example of \textit{Graph Caching Bugs}. In this example, \subject{} incorrectly reuses the cached graph for the last two execution iterations, which leads to incorrect outputs.}
\label{lst:error2_example}
\end{listing}

%% file: 3-2-2-OperatorRelatedError.tex
\subsubsection{Operator-related Bugs.}\label{subsec:sec3_operator_related_error}
Operator-related bugs are the most prevalent type of correctness bugs in \subject{}, accounting for approximately 37.9\% of all cases in our dataset.
These bugs occur in the operator optimization and low-level code generation stages of \subject{}, where high-level DL operators are decomposed, fused, and translated into low-level code for backend compilers such as \texttt{GCC}.
We summarize two steps that are particularly error-prone when compiling operators to low-level code: \textit{operator transformation}, and \textit{low-level code generation}.

\textbf{Bug 3: Operator Transformation Bugs (21/116).}
\textit{Operator Transformation Bugs} refer to correctness bugs arising from incorrect operator transformation.
During the operator transformation process, \subject{} translates high-level DL computational operators (e.g., \texttt{conv}) into low-level computational primitives (e.g., \texttt{matmul}).
Meanwhile, optimizations such as operator fusion may be applied.
Bugs during this transformation could lead to semantic mismatches between low-level primitives and high-level user-defined models, ultimately resulting in wrong model outputs.

\textit{\textbf{Root Cause}}.
We identify three root causes for operator transformation bugs:
\begin{itemize}[leftmargin=*]
    \item \textit{Computational Semantic Mismatch (13/21)}:
    It refers to the mismatch between the original model's computational semantics and its low-level primitive representation after operator transformation.
    This type includes the mismatch of key operator parameters semantics, boundary scenario handling logics (e.g., empty tensors), and the operator computation logics during the transformation.
    \item \textit{Aggressive Optimization and Numerical Instability (6/21)}:
    It includes aggressive optimization decisions and optimization-induced numerical instability.
    Specifically, optimizations (e.g., operator fusion) may lack strict semantic preservation, leading to incorrect optimization decisions (i.e., the optimized graph breaks the original semantics) or amplifying numerical errors for sensitive operators (e.g., exponential, logarithmic), thus resulting in incorrect model outputs. 
    \item \textit{Transformation Incompatibility (2/21)}:
    It refers to the incompatibility between the transformation logic and specific hardware requirements or other dependencies.
    Specifically, the transformation rules may not align with certain hardware architectures or dependencies, leading to wrong model outputs.
\end{itemize}

\textit{\textbf{Bug-triggering Pattern}}.
We identify three bug-triggering patterns that lead to correctness bugs during operator transformation:
\begin{itemize}[leftmargin=*]
    \item \textit{DL Operator with Non-default Parameters or Edge-Case Inputs (11/21)}:
    Bugs are triggered by high-level DL operators (i.e., non-primitive operators) with non-default parameters or edge-case inputs.
    Specifically, 7 out of 11 cases are caused by non-default DL operator parameters, and the remaining 4 are caused by edge-case inputs, which include specific tensor shapes, dtypes, or extreme values.
    \item \textit{Optimization-Triggering DL Operator Sequence (9/21)}:
    Bug-triggering models include specific operator sequences that mainly fall into two categories: numerical computation chains (e.g., \texttt{addmm+cat}, \texttt{matmul+exp}, \texttt{conv+add(alpha)}) and tensor reshaping chains (e.g., \texttt{split+stack+tanh}, \texttt{split+cat}).
    These operator sequences often include non-standard configurations like complex datatype computation.
    \item \textit{Configuration-dependent Scenario (1/21)}:
    This corner case bug is triggered when compiling a DL model on a specific hardware (i.e., \texttt{MPS} device).
\end{itemize}

\textit{\textbf{Example Bug}}.
Listing~\ref{lst:error3_example} is an example of an operator transformation bug~\cite{op_transformation_error}.
In this example, the bug-triggering pattern is a single high-level operator \texttt{torch.diag\_embed} with a specific non-default parameter configuration (i.e., negative \texttt{dim1}).
When compiling this model, \subject{} applies operator decomposition to break down the high-level operator into multiple low-level primitives.
However, this decomposition lacks considering corner cases when the parameter \texttt{dim1} is negative, thus leading to an incorrect \texttt{unsqueeze} operation in the decomposed model.

\input{listings/error_examples/error3.tex}

\begin{tcolorbox}[MyFrame]
\textbf{Summary of Operator Transformation Bugs}:
\textit{Operator Transformation Bugs} accounts for 18.1\% (21/116) of cases in our dataset.
The dominant root cause for this bug type is the inadequate preservation of the original computational semantics.
These bugs are mainly triggered by two patterns: (1) the use of high-level DL operators with non-default parameters or edge-case inputs, and (2) specific operator sequences that trigger aggressive optimizations like operator fusion.
\end{tcolorbox}

\textbf{Bug 4: Low-level Code Generation Bug (23/116).}
\textit{Low-Level Code Generation Bugs} refer to correctness bugs caused when \subject{} translates Python-level DL operators into incorrect low-level code (e.g., C++).
Generating low-level code in \subject{} involves a series of complex hardware-specific transformations and optimizations, including memory alignment and the implementation of hardware-native primitives (e.g., \texttt{triton\_matmul}, \texttt{cuda\_sum}).
Improper low-level code generation can lead to faulty executables with incorrect functionalities or suffer from bugs such as undefined behavior (e.g., non-deterministic tensor values).

\textit{\textbf{Root Cause}}.
The main root cause is the amplified numerical instability~\cite{deepstability}, followed by the lack of semantic preservation during the code generation.

\begin{itemize}[leftmargin=*]
    \item \textit{Numerical Instability \& Precision Mismatch (13/23)}:
    \subject{}'s low-level code generation process may fail to accurately preserve the numerical semantics of high-level PyTorch operators, resulting in amplified numerical errors in the generated code.
    Specifically, this issue manifests in two key forms: the adoption of numerically unstable implementations for computationally sensitive operators (e.g., \texttt{acosh}) and inconsistencies in precision promotion rules between high-level PyTorch operations and the low-level code.
    \item \textit{Lack of Boundary Scenario Handling (7/23)}:
    \subject{} may fail to preserve critical boundary scenarios (e.g., specific parameter bounds) in its generated low-level code, even though these are respected in the implementation of Python-level DL operators.
    This arises because Python-level DL operators (e.g., \texttt{torch.polygamma}) rely on complex logic to handle boundary cases such as \texttt{inf}-valued tensors, which may be incompletely preserved when translated to low-level code.
    \item \textit{Low-Level Language Incompatibility (3/23)}:
    This subcategory arises when the code generation process fails to consider the inherent language features, syntax rules, and type constraints of the target low-level code.
\end{itemize}

\textit{\textbf{Bug-triggering Pattern}}.
We observe that these incorrect low-level code generation bugs are \textbf{exclusively triggered by compiling individual DL operators}, rather than specific operator combinations.
Instead of a specific model structure, the bug-triggering patterns of this bug type are closely related to the input conditions (e.g., tensor values or shapes) and parameter values of the single operator being compiled.
Specifically, we summarize the following two major bug-triggering patterns:
\begin{itemize}[leftmargin=*]
    \item \textit{Extreme Values for Numerical Sensitive Operators (13/23)}:
    Bugs are triggered when compiling numerically sensitive operators (e.g., \texttt{acosh}, \texttt{tanh}) with extreme input values (e.g., large-sized tensors, tensors with extreme values) that amplify floating-point errors in the generated code.
    \item \textit{Single Primitive Operator with Unhandled Boundary Scenarios (10/23)}:
    The compiled operators are specific primitive DL operators (e.g., \texttt{torch.polygamma}) that have unhandled boundary scenarios in their low-level code generation logic, such as specific parameter boundaries or boundary scenario input conditions (e.g., tensors with \texttt{inf} values, scalar tensors).
\end{itemize}

\textit{\textbf{Example Bug}}.
Listing~\ref{lst:error5_example} is an example of a low-level code generation bug in \subject{}~\cite{op_implementation_error}.
In this example, the C++ implementation of the operator \texttt{polygamma} misses the boundary scenario when the parameter \texttt{n} is `1'.
This leads to a less numerically stable implementation of the operator, resulting in larger numerical errors in the compiled model.

\input{listings/error_examples/error5.tex}

\begin{tcolorbox}[MyFrame]
\textbf{Summary of Low-level Code Generation Bugs}:
\textit{Low-level Code Generation Bugs} are the second most prevalent subcategory in our dataset, accounting for 19.8\% (23/116) of all cases.
Most of these bugs are caused by numerical issues in the generated low-level code, such as the use of numerically unstable implementations for sensitive operators and inconsistencies in precision promotion rules.
Triggering patterns for these bugs are exclusively related to compiling a single DL operator with specific input conditions, including extreme values and boundary scenarios.
\end{tcolorbox}

%% file: listings/error_examples/error3.tex
\begin{listing}
\begin{lstlisting}
|\textbf{\underline{User-Defined Model}}|
@torch.compile
def model(x):
    return torch.diag_embed(input=x, dim1=-1,dim2=0,offset=1)
|\textbf{\underline{\textcolor{red}{Buggy Transformed Model}}}|
...
unsqueeze = aten.unsqueeze(cat, |\textbf{\textcolor{red}{6}}|)
permute = aten.permute(unsqueeze, |\textbf{\textcolor{red}{[6, 0, 1, 2, 3, 4, 5]}}|)
...
|\textbf{\underline{\textcolor{codegreen}{Fixed Transformed Model}}}|
...
unsqueeze = aten.unsqueeze(cat, |\textbf{\textcolor{codegreen}{0}}|)
permute = aten.permute(unsqueeze, |\textbf{\textcolor{codegreen}{[0, 1, 2, 3, 4, 5, 6]}}|)
...
\end{lstlisting}
\caption{Example of \textit{Operator Transformation Bugs}. In this example, \subject{} incorrectly decomposes the \texttt{torch.diag\_embed} operator into a sequence of operators with incorrect parameters, which leads to incorrect outputs.}
\label{lst:error3_example}
\end{listing}

%% file: listings/error_examples/error5.tex
\begin{listing}
\begin{lstlisting}
|\textbf{\underline{User-Defined Model}}|
@torch.compile
def model(x):
    return torch.special.polygamma(x, n=1)
|\textbf{\underline{\textcolor{red}{Buggy Op Implementation}}}|
lambda x, y: f"{x} == 0 ? calc_digamma({y}) : calc_polygamma({y}, {x})",
|\textbf{\underline{\textcolor{codegreen}{Fixed Op Implementation}}}|
lambda x, y: f"{x} == 0 ? calc_digamma({y}) : ({x} == 1 ? trigamma({y}) : calc_polygamma({y}, {x}))",
\end{lstlisting}
\caption{Example of \textit{Low-Level Code Generation Bugs}. In this example, \subject{} generates incorrect C++ code for the \texttt{polygamma} operator by missing the special case when $n=1$, which leads to incorrect outputs.}
\label{lst:error5_example}
\end{listing}

%% file: 3-2-3-MemoryRelatedError.tex
\subsubsection{Memory-related Bugs.}\label{subsec:sec3_memory_related_error}
Memory-related bugs are the second most prevalent bug type in our dataset, accounting for 33.6\% of all cases.
These bugs refer to the incorrect memory management in \subject{}, which includes two subcategories: \textit{in-place operation handling bugs} and \textit{memory layout conflicts}.

\textbf{Bug 5: In-Place Operation Handling Bugs (25/116).}
\textit{In-Place Operation Handling Bugs} refer to correctness bugs caused when \subject{} incorrectly optimizes or tracks in-place operations in user-defined models.
User-defined DL models often incorporate in-place operations (e.g., \texttt{torch.sin\_}) that directly mutate tensor memory to avoid redundant tensor cloning/copying and reduce memory overhead.
Mishandling of these operations can result in incorrect memory reads/writes and, consequently, wrong outputs of compiled models.
Notably, this bug type is the \textbf{most prevalent} subcategory, accounting for 21.6\% (25/116) in our dataset.

\textit{\textbf{Root Cause}}.
The root cause of this bug type can be subdivided into the following two categories:
\begin{itemize}[leftmargin=*]
    \item \textit{Incorrect Alias Tracking (17/25)}:
    \subject{} misses or misconstructs aliasing relationships among variables such as tensors used in in-place operations.
    This incorrect aliasing may misdirect tensor mutations (e.g., in-place operations) to unintended memory regions.
    For example, when \texttt{unfold} creates a memory-overlapping view \texttt{y} of \texttt{x}, \subject{} fails to track this aliasing relationship between \texttt{y} and \texttt{x}.
    As a result, the in-place operation \texttt{y.abs\_()} only modifies the elements of \texttt{y}, but does not propagate the modification to \texttt{x}, which leads to incorrect model outputs.
    \item \textit{Incorrect In-Place Operation Optimizations (8/25)}:
    \subject{} produces wrong computational graphs when applying in-place operation-specific optimizations, such as functionalization~\cite{functionalization}-a transformation that rewrites in-place operations as out-of-place ones for further optimization.
    For instance, one bug in this subcategory causes an operator \texttt{as\_strided} to be incorrectly eliminated when \subject{} is performing functionalization on an in-place operation \texttt{resize\_}.
\end{itemize}

\textit{\textbf{Bug-triggering Pattern}}.
Although correctness bugs in this subcategory are caused by incorrect handling of in-place operations, we find that simply executing in-place operations is not sufficient to trigger this bug type.
Instead, a common pattern that triggers this bug type is \textbf{applying in-place operations on tensor aliases}, which are often created by specific operations (e.g., \texttt{torch.Tensor.view}, \texttt{torch.Tensor.expand}).
This pattern commonly involves two operations: one for creating the aliasing relationships and one for applying in-place mutation.

\textit{Creation of Aliasing Relationships}:
When triggering the incorrect in-place operation handling bug, the aliasing relationships between tensors are often created through explicit tensor view operations (e.g., \texttt{reshape} and \texttt{permute}) and slicing (e.g., basic/strided/masked indexing, offset slicing).
Beyond these explicit tensor aliasing, implicit aliasing is also a common pattern that leads to this bug type, which includes aliases created through customized operators, nested mutable Python objects (e.g., lists/dictionaries containing tensors), and tensor wrapping (e.g., using \texttt{torch.nn.Parameter}).

\textit{Applied In-place Operations}:
The dominant type of in-place operations that trigger this bug type is value assignment, which includes indexed/masked tensor updates (e.g., \texttt{out[mask1] *= 0.5}) and arithmetic/numerical operations (e.g., \texttt{add\_}, \texttt{clamp\_}, \texttt{abs\_}).
Other types of in-place operations that trigger this bug type include shape/storage modifications (e.g., \texttt{resize\_}, \texttt{a.data = b}) and implicit mutations (e.g., Triton \texttt{tl.store}, nonlocal variable assignment).

\textit{\textbf{Example Bug}}.
Listing~\ref{lst:error6_example} is an example of \textit{In-Place Operation Handling Bugs}~\cite{inplace_op_error}.
In this example, the bug-triggering model first creates an alias of the input tensor \texttt{x} using the \texttt{expand} operation, then performs an in-place modification on the aliased tensor.
However, \subject{} fails to correctly propagate the in-place mutation, resulting in incorrect modifications on elements except for the first one.
\input{listings/error_examples/error6.tex}

\begin{tcolorbox}[MyFrame]
\textbf{Summary of In-Place Operation Handling Bugs}:
\textit{In-Place Operation Handling Bugs} are the most prevalent type, accounting for 21.6\% (25/116) in our dataset.
These bugs are primarily caused by (1) the incorrect tracking of aliasing relationships between tensors used by in-place operations and (2) the wrong graph produced when optimizing in-place operations.
Our analysis reveals that covering the in-place operations alone is not sufficient to trigger this bug type; instead, the common bug-triggering pattern is to apply these operations on tensor aliases created by specific operations (e.g., \texttt{torch.Tensor.view}, \texttt{torch.Tensor.expand}).
\end{tcolorbox}

\textbf{Bug 6: Memory Layout Conflicts (14/116).}
\textit{Memory Layout Conflicts} describe mismatches between tensor memory layouts (e.g., contiguous vs.\ non-contiguous, channel-first vs.\ channel-last) and the layout constraints of downstream operators during compilation.
Such mismatches lead to invalid tensor data access and interpretation, resulting in incorrect outputs of the compiled model.

\textit{\textbf{Root Cause}}.
We identify two root causes for this bug type:

\begin{itemize}[leftmargin=*]
    \item \textit{Incorrect Tracking of Layout Metadata (6/14)}:
    It refers to the incorrect preservation or propagation of tensor memory layout metadata (e.g., tensor contiguity, stride values) during compilation.
    This type of bug often occurs when \subject{} mutates tensor layouts without proper metadata synchronization.
    \item \textit{Missing Compatibility Checks for Layout Transformations (8/14)}:
    \subject{} applies layout transformations (e.g., channel-last conversion, tensor stride mutation) without verifying if the new metadata aligns with the requirements of downstream operators and devices.
    This type of bug is caused by the incomplete layout constraint capture in the compiler for certain operators (e.g., the \texttt{SDPA} operator, custom Triton kernels).
\end{itemize}

\textit{\textbf{Bug-triggering Pattern}}.
User-defined models that trigger this bug type typically contain specific operator sequences that mutate the tensor memory layout.
Specifically, we summarize the common categories of bug-triggering patterns.

\begin{itemize}[leftmargin=*]
    \item \textit{Optimization-Driven Layout Transformations (7/14)}:
    The user-defined models include operator sequences that trigger the internal optimization (e.g., operator fusion, stride reordering) in \subject{}, which implicitly alters tensor layout attributes like stride values or contiguity.
    These optimizations fail to propagate updated layout metadata to subsequent operations, leading to mismatches between metadata and actual memory storage.
    \item \textit{Operator-Induced Layout Transformations (7/14)}:
    The user-defined models involve operators that directly modify tensor layout (e.g., transpose, in-place ops) or rely on specific layout constraints (e.g., requiring contiguous tensors in specific dimensions).
    These operators either fail to adapt to non-standard layouts (e.g., tensor in non-contiguous layout) or include strict layout constraints, causing mismatches between layout metadata and memory storage.
\end{itemize}

\textit{\textbf{Example Bug}}.
Listing~\ref{lst:error7_example} demonstrates a memory layout conflict bug~\cite{memory_layout_conflict_error}.
In this case, a specific DL operator (\texttt{torch.argwhere}) produces a tensor with a non-contiguous memory layout, which is incompatible with the subsequent indexing operation, and \subject{} fails to handle this layout incompatibility.
As a result, the low-level \texttt{C++} code contains incorrect memory access instructions.

\input{listings/error_examples/error7.tex}

\begin{tcolorbox}[MyFrame]
\textbf{Summary of Memory Layout Conflicts}:
\textit{Memory Layout Conflicts} refer to the mismatches between the actual tensor memory layout and the layout constraints of downstream operators.
These bugs are primarily caused by the incorrect tracking of layout metadata and the missing/lack of compatibility checks for layout transformations.
The common bug-triggering pattern for this bug type is to induce memory layout transformations through specific operator sequences.
\end{tcolorbox}

%% file: listings/error_examples/error6.tex
\begin{listing}
\begin{lstlisting}
|\textbf{\underline{User-Defined Model}}|
@torch.compile
def model(x):
    y = x.expand(2, *x.shape)
    y[0, 0] = 5
    return y
x = torch.tensor([0,1,2,3,4,5])
model(x)
|\textbf{\underline{\textcolor{red}{Buggy Output}}}|
print(x)  # |\textbf{\textcolor{red}{tensor([5,2,4,6,8,10]) -- Incorrect, elements except the first should not be mutated}}|
|\textbf{\underline{\textcolor{codegreen}{Fixed Output}}}|
print(x)  # |\textbf{\textcolor{codegreen}{tensor([5,1,2,3,4,5]) -- Correct}}|
\end{lstlisting}
\caption{Example of \textit{In-Place Operation Handling Bugs}. In this example, \subject{} fails to handle the in-place mutation of the expanded tensor correctly, which leads to unintended mutations on the original input tensor.}
\label{lst:error6_example}
\end{listing}

%% file: listings/error_examples/error7.tex
\begin{listing}
\begin{lstlisting}
|\textbf{\underline{User-Defined Model}}|
@torch.compile
@torch._dynamo.config.patch(capture_dynamic_output_shape_ops=True)
def model(tensor, mapping,):
    xx, yy = torch.meshgrid(mapping, tensor, indexing="ij")
    indices = torch.argwhere(xx == yy)  # indices.stride = (1, 3)
    mapped_values = torch.zeros_like(tensor)
    mapped_values[indices[:, 1]] = indices[:, 0]  # |\textbf{\textcolor{red}{`index\_put' assumes contiguous layout.}}|
    return mapped_values

|\textbf{\underline{Low-Level C++ code}}|
for(int64_t x0=static_cast<int64_t>(0L); x0<static_cast<int64_t>(ks0); x0+=static_cast<int64_t>(8L))
{
    ...
    |\textbf{\underline{\textcolor{red}{- ... = loadu(in\_ptr0 + int64\_t(1L + x0), ...); // Buggy Low-Level IR}}}|
    |\textbf{\underline{\textcolor{codegreen}{+ ... = loadu(in\_ptr0 + int64\_t(ks0 + x0), ...); // Fixed Low-Level IR}}}|
    ...
}
\end{lstlisting}
\caption{Example of \textit{Memory Layout Conflicts}. In this example, \subject{} fails to consider the memory layout of the `indices' when generating C++ code for `index\_put', leading to incorrect memory access.}
\label{lst:error7_example}
\end{listing}

%% file: 5-Benchmark.tex
\section{Effectiveness of Existing Testing Techniques}\label{sec:benchmark}
This section evaluates the effectiveness of existing DL compiler testing techniques in detecting correctness bugs in \subject{}.
We aim to answer the following research questions:
\begin{itemize}[leftmargin=*,topsep=0pt]
    \item \textbf{RQ1:} How effective are existing DL compiler testing techniques in detecting real-world correctness bugs in \subject{}?
    \item \textbf{RQ2:} What are the strengths and weaknesses of existing DL compiler testing techniques in detecting correctness bugs in \subject{}?
\end{itemize}

\subsection{Selection and Configuration of DL Compiler Testing Techniques}
We select and configure the five latest DL compiler testing techniques for evaluation, including two grammar-based techniques, one test migration technique, and two LLM-based techniques, as shown in Table~\ref{tab:fuzzer_characteristics}.
Following the common practice in DL compiler testing evaluation~\cite{whitefox,nnsmith,neuri}, each technique is configured with its default settings.

\noindent{}\textbf{NNSmith}~\cite{nnsmith}:
NNSmith is a grammar-based fuzzer that generates DL models with pre-defined constraints.
We directly execute NNSmith using the fuzzing script provided by its authors, specifying the backend as `pt2' for fuzzing \subject{}.
After fuzzer execution, we collect all bug-revealing test cases reported by NNSmith and utilize its provided API `nnsmith.report\_syn' to synthesize executable bug reports.

\noindent{}\textbf{NeuRI}~\cite{neuri}:
NeuRI is another grammar-based fuzzer, generating DL models via inductive inference operator constraints.
We run NeuRI with its default configurations, setting the backend to `torchcomp' for \subject{} fuzzing.
Notably, we observe that NeuRI only outputs synthesized models and their corresponding test inputs, rather than integrated test cases.
To address this, we develop a custom script to synthesize integrated, executable test cases with NeuRI-generated models and inputs\@.

\noindent{}\textbf{Opera}~\cite{opera}:
Opera is a test migration tool that collects test suites from DL libraries for DL compiler fuzzing.
For \subject{} fuzzing, we directly utilize all PyTorch test cases collected by Opera, which include predefined model structures and corresponding inputs.
We further develop a simple script to integrate these components into executable test cases.

\noindent{}\textbf{WhiteFox}~\cite{whitefox}:
WhiteFox is an LLM-based fuzzer that leverages LLMs to analyze optimization code and generate optimization-aware test cases.
We directly adopt the test cases published by their authors, which were generated specifically for fuzzing \subject{}—achieving the best performance reported in their paper.

\noindent{}\textbf{DeepConstr}~\cite{deepconstr}:
DeepConstr is another LLM-based fuzzer, which leverages LLMs to extract complete constraints for fuzzing \subject{}.
To run DeepConstr, we use the officially provided fuzzing command, ensuring the backend type, filter type, and the `mgen' configurations are aligned with the artifact.

\input{tables/Fuzzer-Characteristics.tex}

\subsection{Benchmark Construction}
We construct a benchmark from our dataset collected in Section~\ref{sec:empirical_study} to evaluate the correctness bug detection capabilities of the studied techniques.
To determine whether a bug has been detected by a technique, existing studies~\cite{apifuzzing_benchmarking,fuzz_testing_benchmarking} typically rely on execution symptoms (e.g., error messages, crashes) as well as manual inspection.
However, correctness bugs in \subject{} do not raise exceptions, crashes, or warnings, making it difficult to verify their detection solely from execution symptoms.
To address this issue, we instead leverage developers' patches to judge whether a technique has detected a specified correctness bug.
Specifically, we adopt the following detection criteria:
\textit{A technique $F$ is considered to detect the specified correctness bug $E$ if any bug reported by $F$ is fixed by the patch of $E$.}

Formally, let $V_B$ and $V_F$ be the buggy and fixed versions of \subject{} before and after applying the patch of $E$, respectively.
We determine whether technique $F$ detects the specified correctness bug $E$ using the following rule:

\begin{equation*}
    \exists~S_F \in S_F^{all}: \text{Error}(S_F,V_B) \land \neg \text{Error}(S_F,V_F)
\end{equation*}

where $S_F^{all}$ is the set of all bug-revealing test cases reported by $F$ and $\text{Error}(S_F,V)$ is a function that checks whether the bug-revealing test case $S_F$ can trigger a correctness bug in version $V$.

Accordingly, our benchmark only includes correctness bugs with clear developer patches and can be reproduced in the public release version of \subject{} without specialized hardware (beyond CPU/GPU) or platform constraints.
In total, 77 correctness bugs are included in our benchmark after systematic filtering (as detailed in Table~\ref{tab:sec4_benchmark_construction}).

\input{tables/Benchmark-Collection.tex}

\subsection{Experiment Setup}
We run each testing technique across released versions of PyTorch (from 2.0 to 2.7), covering the affected versions for all 77 correctness bugs in our benchmark.
After executing each technique, we gather all test cases that trigger correctness bugs and apply our detection criteria to measure each technique's effectiveness in detecting correctness bugs.
Specifically, we ran all bug-revealing test cases reported by each testing technique on both the buggy and fixed versions of \subject{}, for each correctness bug in our benchmark.
The technique is deemed to detect a correctness bug if any reported test case meets the detection criterion (i.e., can trigger the bug in the buggy version but not in the fixed version).
After this automatic measurement, we manually verify the triggering test cases for each detected correctness bug, confirming that all claimed detections are valid.

Note that our assessment only evaluates whether each testing technique can detect the correctness bugs in our benchmark.
We exclude bugs with other symptoms (e.g., program crashes) as they are not the focus of our study.
We also don't report correctness bugs in \subject{} outside our benchmark, because it is difficult to determine the uniqueness of these bugs without developer patches.
Therefore, our results do not reflect the overall bug detection performance of these techniques.

\noindent{}\textbf{Environment.}
The execution environment for the studied techniques is a 32-core server with three 3090Ti GPUs\@.

\subsection{RQ1: Effectiveness in Detecting Real-World Correctness Bugs}

\begin{figure}[t!]
    \centering
    \includegraphics[width=0.6\linewidth]{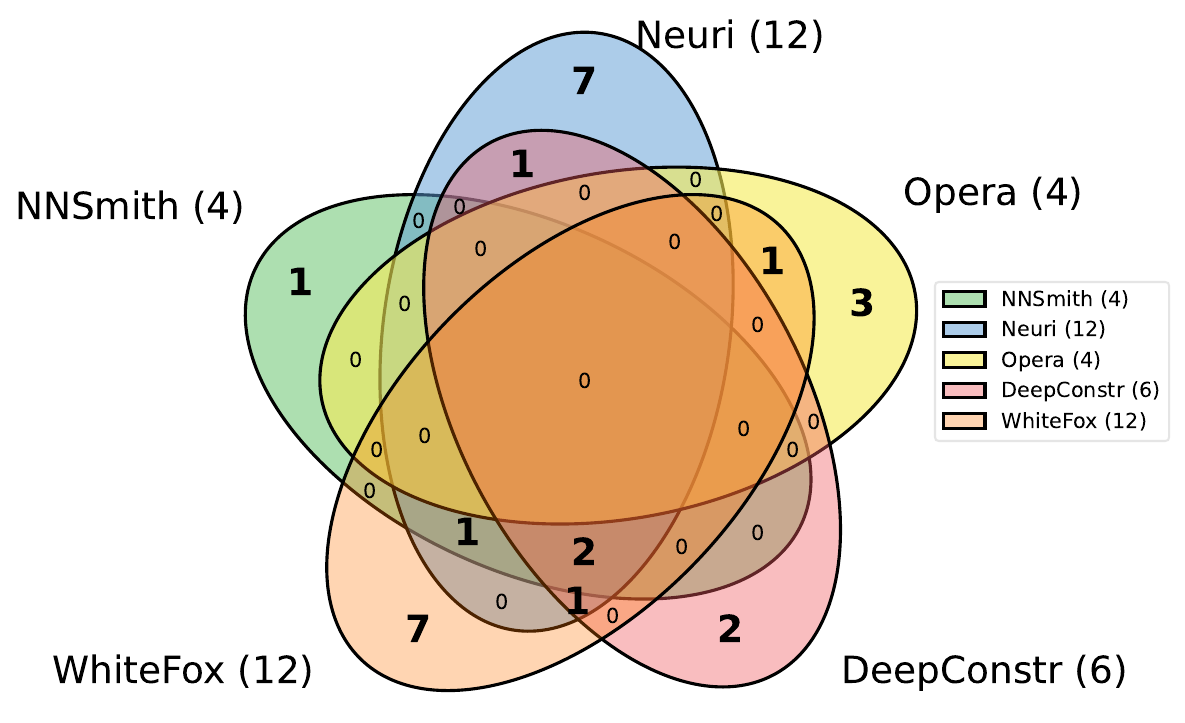}
    \caption{Benchmarking studied techniques' bug detection performance}\label{fig:benchmark_bug_detection_venn}
    \Description{Benchmarking Existing Techniques' Bug Detection Performance}
\end{figure}

Figure~\ref{fig:benchmark_bug_detection_venn} presents the correctness bug detection effectiveness of the studied techniques.
These techniques collectively detect 26 out of the 77 correctness bugs in our benchmark, highlighting the significant room for improvement.
Specifically, NeuRI~\cite{neuri} and WhiteFox~\cite{whitefox} achieve the best performance by detecting 12 correctness bugs each, followed by DeepConstr~\cite{deepconstr} (6 detected), Opera~\cite{opera} (4 detected), and NNSmith~\cite{nnsmith} (4 detected).
Notably, all techniques complement each other in correctness bug detection, as each detects unique bugs.
For instance, 7 of the 12 correctness bugs detected by NeuRI and WhiteFox are not identified by any other technique.

\begin{figure}[t!]
    \centering
    \includegraphics[width=\linewidth]{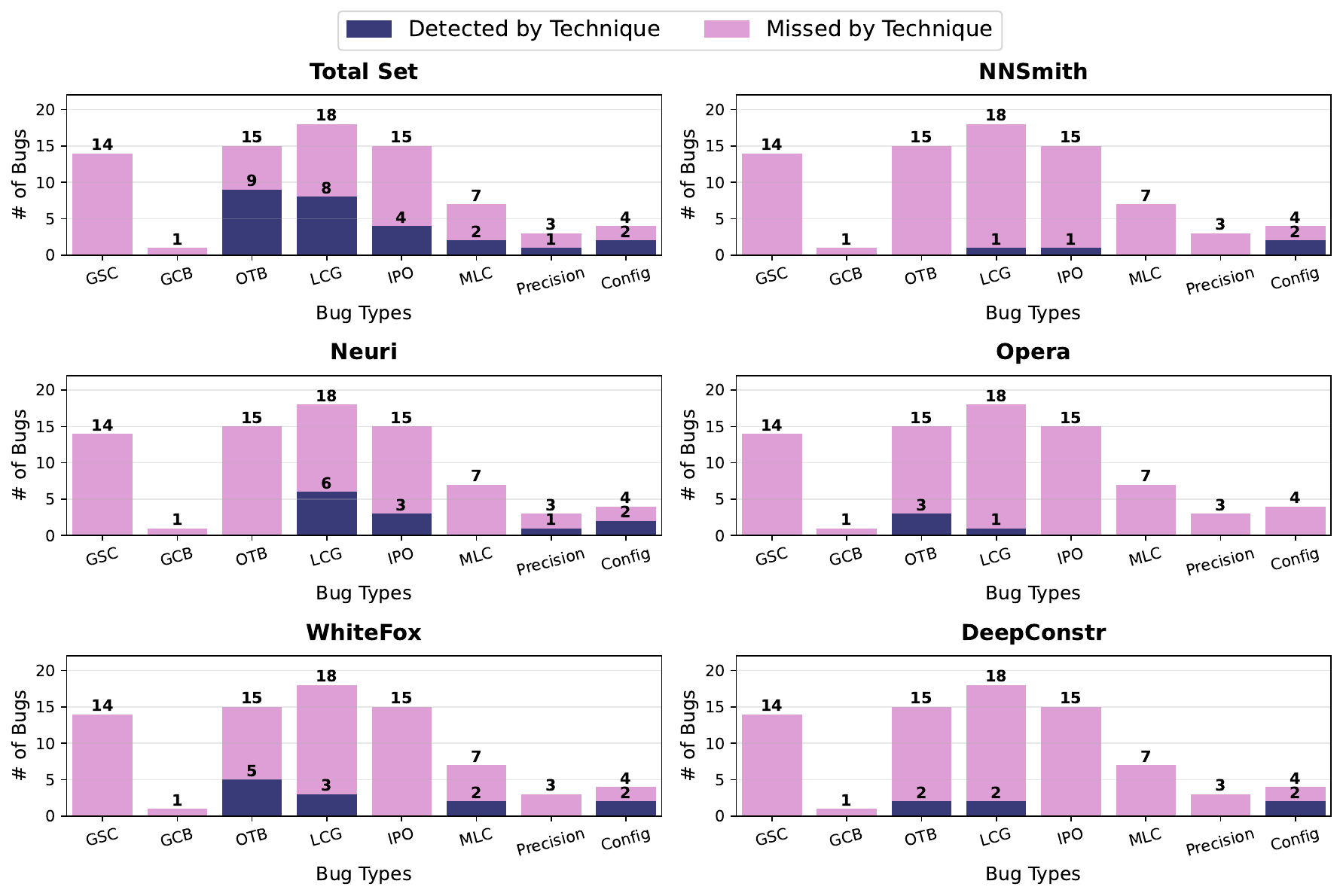}
    \caption{Distribution of detected and missed bugs. The name of each bug category is abbreviated using its initials (e.g., `OTB' for \textit{Operator-Transformation Bugs}, `LCG' for \textit{Low-Level Code Generation Bugs}).}\label{fig:sec5_missed_bugs_all_fuzzers}
    \Description{Distribution of Detected and Missed Bugs by All Fuzzers}
\end{figure}

Based on the bug type categorization proposed in Section~\ref{sec:empirical_study}, we further analyze the distribution of detected and missed correctness bugs (see Figure~\ref{fig:sec5_missed_bugs_all_fuzzers}).
Among the 26 cases detected by the studied techniques, operator-related bugs (including `OTB' and `LCG') are the most frequently detected type (17/26), followed by memory-related bugs (including `IPO' and `MLC') (6/26).
It's worth noting that none of the graph-related bugs (including `GSC' and `GCB') in our benchmark are detected by studied techniques, revealing a critical gap in \subject{} technique designing and enhancement.
Regarding the correctness bugs missed by all studied techniques, 16 of the operator-related bugs in our benchmark are missed, 16 out of 22 memory-related bugs are missed, and all graph-related bugs remain undetected.

\begin{tcolorbox}[MyFrame]
\textbf{Findings and Advice from RQ1}:
Studied techniques are limited in detecting correctness bugs in \subject{}, with only 26 out of 77 cases in our benchmark detected.
Within the detected cases, operator-related bugs are the most frequently detected type, while a significant portion of memory-related bugs and all graph-related bugs in our benchmark are missed by the studied techniques.
To address these gaps, future techniques could prioritize detecting graph- and memory-related bugs.
\end{tcolorbox}

\subsection{RQ2: Strengths and Weaknesses in Detecting Correctness Bugs}

This research question investigates the characteristics of correctness bugs detected and missed by the studied techniques in our benchmark, based on the bug type categorization in Section~\ref{sec:empirical_study}.

\subsubsection{Characteristics of missed graph-related bugs}
All 15 graph-related bugs in our benchmark are missed by studied techniques, including 14 \textit{Graph Semantic Capturing Bugs} (GSC) and 1 \textit{Graph Caching Bug} (GCB).
By analyzing the bug-triggering patterns of these missed graph-related bugs, we identify two reasons for their omission.

Firstly, \textbf{testing oracle limitations} prevent studied techniques from detecting a subset of graph-related bugs.
Most current DL compiler techniques rely on differential testing as a test oracle, which executes the compiled model once and compares model outputs before and after compilation.
However, as highlighted in Section~\ref{sec:empirical_study}, certain graph-related bugs can only be detected by dedicated testing pipelines (i.e., repeated execution), rather than standard differential testing.
For instance, one graph-related bug in our benchmark requires repeated runs of the compiled model with distinct input values to detect abnormal consistency in outputs.
Overall, 4 of the 15 missed graph-related bugs require customized test oracles and thus remain undetected by the studied techniques.

Secondly, \textbf{insufficient API coverage} in the model structure is another reason for missing graph-related bugs.
While current testing techniques focus on generating models with diverse DL computational operators, they do not cover other critical API categories, such as non-computational APIs (e.g., control-flow APIs) and context-related APIs (e.g., \texttt{TorchDispatchMode}).
In particular, 6 missed graph-related bugs are associated with non-computational operators, and another 5 arise from execution context mutation.
Indeed, we observe that these critical APIs are rarely included in their generated models, which explains the reason why these graph-related bugs are missed.

\begin{tcolorbox}[MyFrame]
\textbf{Findings and Advice for Graph-Related Bug Detection}:
We identify two key reasons for studied techniques' ineffectiveness in detecting graph-related bugs in our benchmark: (1) testing oracle limitations, as some graph-related bugs require dedicated testing pipelines rather than standard differential testing; (2) insufficient API coverage in generated models, as studied techniques lack covering specific APIs critical for triggering graph-related bugs.
To improve the detection of graph-related bugs, future techniques should develop customized testing pipelines and oracles tailored for graph-structure bugs, and enhance model generation to cover a richer set of APIs, including context-related and non-computational APIs.
\end{tcolorbox}

\subsubsection{Characteristics of detected and missed operator-related bugs}
Among all 33 operator-related bugs in our benchmark, 16 are successfully detected by the studied testing techniques, including 9 \textit{operator transformation bugs (OTBs)} and 8 \textit{low-level code generation bugs (LCGs)}.

Examining the detected bugs, we observed that \textbf{most studied techniques are more effective at catching single-operator semantic mismatch bugs than at detecting optimization bugs that require specific operator sequences to trigger}.
Specifically, the majority (14 out of 17) of these detected bugs are caused by semantic mismatches when compiling a single DL operator with specific parameters or edge-case inputs, covering 6 OTBs and all 8 detected LCGs.
The remaining 3 optimization-related bugs, which are exclusively detected by WhiteFox~\cite{whitefox}, are triggered by operator sequences that invoke optimization passes such as operator fusion.
This unique detection capability stems from WhiteFox's white-box static analysis, which mines optimization-triggering operator sequence patterns to guide test generation.

For the remaining 16 missed bugs (6 OTBs and 10 LCGs), we identify two key properties that explain detection failures.
Firstly, 13 of the 16 missed bugs are single-operator semantic mismatch bugs, the same class that the studied techniques are generally effective at detecting.
These operators include numerically sensitive operators (e.g., \texttt{acosh}, \texttt{gumbel\_softmax}), domain-specific DL operators (e.g., \texttt{pad}, \texttt{polygamma}), and high-level operators with complex parameter spaces (e.g., \texttt{BatchNorm2d}, \texttt{flex\_attention}).
Although all these bug-triggering operators are covered by at least one studied technique, they still fail to trigger these bugs, suggesting that operator coverage alone is insufficient for bug detection.
Indeed, all these missed single-operator bugs require specific input patterns, such as boundary parameter values and extreme numerical values.
Secondly, the remaining 3 missed bugs are optimization-related and triggered only by specific operator sequences (e.g., {\texttt{unsqueeze} followed by \texttt{mul-neg-div}}).
We observe that these sequences are not systematically generated by the studied techniques, which explains why these optimization-related bugs are missed.

\begin{tcolorbox}[MyFrame]
\textbf{Findings and Advice for Operator-Related Bug Detection}:
We find that most of our studied techniques are more capable of detecting single-operator semantic mismatch bugs than optimization-related bugs, which require specific operator sequences to be triggered.
Even for single-operator semantic mismatches, many bugs are missed despite the target operators being covered by the studied techniques.
To improve the operator-related bug detection effectiveness, future methods should adopt optimization-aware testing strategies that generate specific operator sequences to trigger optimization-related bugs.
Besides, we advise future techniques to not only target operator coverage but also target covering diverse input/parameter patterns, such as boundary scenarios and extreme numerical values.
\end{tcolorbox}

\subsubsection{Characteristics of detected and missing memory-related bugs}
Studied techniques collectively detect 6 out of the 22 memory-related bugs in our benchmark, including 4 \textit{in-place operation handling bugs (IPOs)} and 2 \textit{memory layout conflicts (MLCs)}.
We further characterize the features of the detected and missed bugs.

The two detected MLC bugs, both identified by WhiteFox~\cite{whitefox}, are triggered by the \texttt{conv2d} operator, which implicitly mutates the memory layout of input tensors and leads to layout incompatibility across the model structure; their root cause falls into the category of \textit{Missing Compatibility Checks for Layout Transformations}.
For the four detected IPO bugs, three are detected by NeuRI~\cite{neuri} and one by NNSmith~\cite{nnsmith}.
All these IPO bugs are triggered by in-place operations on tensor aliases or views created via basic, commonly used alias-generating operators (including slicing, reshape, tile, and zero-padding pad), with a unified root cause of \textit{Incorrect Alias Tracking} during compilation.

For the 16 missed memory-related bugs (including 11 IPO bugs and 5 MLC bugs), our analysis yields a key insight: these bugs are not missed due to insufficient operator coverage for in-place operations, but rather because \textbf{these testing techniques fail to synthesize the complex, bug-triggering usage patterns of these operators.}
To elaborate on this finding, we further break down these missed bugs by their subcategory and analyze their respective bug-triggering patterns.
Among the 11 missed IPO bugs, all in-place operations (except for one customized Triton operator) are already covered by our studied techniques, yet the corresponding bugs remain undetected.
This is because IPO bugs require in-place operations to be applied on complex tensor aliases or views constructed via under-explored patterns (including unfold, nested list indexing, \texttt{.data} access), which studied techniques do not systematically generate.
For the remaining 5 missed MLC bugs, all layout-mutating operators (except for one customized operator) are covered by the studied techniques.
These bugs are still missed because detecting them requires specific model structure configurations to expose the memory layout incompatibility, a critical pattern that studied techniques do not explicitly target.

\begin{tcolorbox}[MyFrame]
\textbf{Findings and Advice for Memory-Related Bug Detection}:
We find that existing testing techniques lack systematic support for complex tensor alias/view patterns in in-place operations and memory layout mutation, which are essential to trigger memory-related bugs.
Thus, a large number of such bugs in our benchmark remain undetected.
To bridge this gap, future methods should incorporate diverse in-place operations on complex tensor aliases/views and richer memory layout mutation patterns in test case generation.
\end{tcolorbox}

\subsubsection{Other Bugs}
We notice that the studied techniques also detect three other types of correctness bugs, including one precision bug~\cite{detected_precision_error} and two configuration bugs~\cite{detected_config_error,detected_config_error2}.

The two detected configuration bugs are caused by the same root cause: incorrect \texttt{GCC} compilation flag used by \subject{} that leads to numerical instability in the compiled executables.
These two bugs are triggered by test cases with similar patterns, including matrix multiplication (i.e., \texttt{matmul} operator) and the softmax operator on large-valued tensors.

The precision bug is detected by specific DL models, including a matrix multiplication and a tangent operation. 
The root cause is that \subject{} incorrectly converts the intermediate variable's data type to \texttt{float16} during compilation, resulting in a critical precision loss in the final output.

%% file: tables/Fuzzer-Characteristics.tex
\begin{table}[htbp]
\centering
\caption{Characteristics of DL compiler testing techniques used in this study.}\label{tab:fuzzer_characteristics}
\renewcommand{\arraystretch}{1.1}
\resizebox{0.5\linewidth}{!}{
\begin{tabular}{l|c|c}
\toprule
\multirow{1}{*}{\textbf{Technique}} & \multirow{1}{*}{\textbf{Year}} & \multirow{1}{*}{\textbf{Generation Strategy}} \\
\midrule
Opera & 2025 & Test Migration  \\
\midrule
DeepConstr & 2024 & LLM-based generation  \\
\midrule
WhiteFox & 2024 & LLM-based generation  \\
\midrule
NeuRI & 2023 & Grammar-based generation  \\
\midrule
NNSmith & 2022 & Grammar-based generation  \\
\bottomrule
\end{tabular}
}
\end{table}

%% file: tables/Benchmark-Collection.tex
\begin{table}[htbp]
\centering
\caption{Step-by-step filtering process for correctness bug benchmark construction}\label{tab:sec4_benchmark_construction}
\renewcommand{\arraystretch}{1.1}
\resizebox{0.7\linewidth}{!}{
\begin{tabular}{l|l|r|r}
\toprule
\textbf{Step} & \textbf{Filtering Criterion / Category} & \textbf{\# Excluded} & \textbf{\# Remaining} \\
\midrule
--- & Studied Dataset (Confirmed Correctness Bugs) & --- & 116 \\
\hline
1 & Excluded: no reproducible code & 6 & 110 \\
2 & Excluded: not torch.compile bug & 1 & 109 \\
3 & Excluded: patch not identified & 17 & 92 \\
4 & Excluded: not reproducible on release & 7 & 85 \\
5 & Excluded: require non-CPU/GPU hardware & 8 & 77 \\
\midrule
\rowcolor{gray!20}
& \textbf{Benchmark} & --- & \textbf{77} \\
\bottomrule
\end{tabular}
}
\end{table}

%% file: 6-Evaluation.tex
\section{AlignGuard: Knowledge-Guided Correctness Bug Detection Tool}\label{sec:proof_of_concept}
In this section, we demonstrate the usefulness of our empirical evidence and suggest future improvement directions by presenting \toolname{}—a preliminary proof-of-concept testing technique tailored to detect correctness bugs in \subject{}.
Following the design of LLM-guided test mutation strategies~\cite{mut4all,issuemut,fuzzgpt}, \toolname{} leverages an LLM to automatically (1) extract historical bug-triggering patterns from our dataset and (2) apply explicit mutations on existing test cases aimed at generating correctness bug-revealing test cases.
The key insight of \toolname{} is that it goes beyond relying solely on historical bug records (e.g., issue reports and fix commits); instead, it combines bug characteristics (including bug categories, bug-triggering patterns, and root causes) from our empirical study with a multi-step mutation workflow to trigger correctness bugs in \subject{}.

\subsection{Design Overview}
\begin{figure}[t!]
    \centering
    \includegraphics[width=\linewidth]{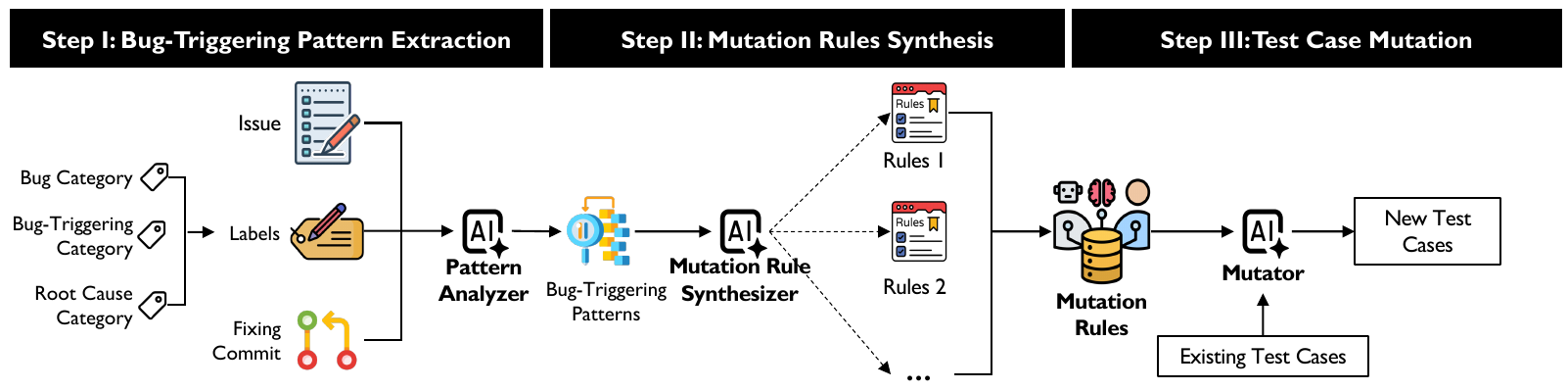}
    \caption{Workflow of \toolname{}}\label{fig:tool}
    \Description{Workflow of \toolname{}}
\end{figure}

Figure~\ref{fig:tool} shows the workflow of \toolname{}.
For each historical correctness bug collected in our empirical study, \toolname{} applies the following three steps to generate test cases:

\textit{Step 1: Bug-Triggering Patterns Extraction (Section~\ref{subsec:tool_pattern_extraction}).}
\toolname{} first employs an LLM to extract historical bug-triggering patterns from the issue report and the fixing commit of the given correctness bug.
Enhanced with the chain-of-thought (CoT) prompting and structured bug characteristics, including bug categories, bug-triggering patterns, and root causes, the LLM outputs explicit, well-formatted bug-triggering patterns for the following mutation rule synthesis.

\textit{Step 2: Mutation Rules Synthesis (Section~\ref{subsec:tool_mutation_synthesis}).}
Based on the extracted patterns, \toolname{} uses an LLM to synthesize dedicated mutation rules. These rules are standardized as a set of <conditions, actions> pairs, which explicitly define applicable scenarios to apply the mutation (i.e., <conditions>) and the corresponding mutation actions to be taken (i.e., <actions>).

\textit{Step 3: Test Case Mutation (Section~\ref{subsec:tool_test_generation}).}
Finally, \toolname{} leverages an additional LLM to generate new test cases by applying the synthesized mutation rules to existing test cases.
These test cases include necessary components for bug detection, including DL model structure, input data, compilation arguments, and the test oracle.

\subsection{General Format of PyTorch Compiler Test Case}\label{subsec:tool_test_format}
To generate structured bug-triggering patterns and actionable mutation rules, we first define the general format of a PyTorch compiler test case used by \toolname{}, which typically comprises four core components:
\begin{itemize}
    \item \textbf{Model Structure}: The architecture of the DL model compiled by \subject{}, usually implemented as a subclass of \texttt{torch.nn.Module} in PyTorch.
    \item \textbf{Input Data}: Inputs conforming to the model's input signature, commonly including tensors and other auxiliary parameter values.
    \item \textbf{Compilation Arguments}: Configuration parameters passed to the compiler, such as optimization levels, target hardware devices, and runtime environment settings.
    \item \textbf{Testing Pipeline}: The end-to-end workflow for validating the output correctness of the compiled model.
    This workflow covers model compilation, execution, and output correctness checking.
    A widely used testing pipeline is a differential testing-based mechanism: compiling the target model and comparing its outputs against those of the eager (uncompiled) model to verify computational consistency.
\end{itemize}

During pattern extraction and mutation rule synthesis, we instruct the LLM to handle each component separately, resulting in more fine-grained and structured outputs.

\subsection{Bug-Triggering Patterns Extraction}\label{subsec:tool_pattern_extraction}
In bug-triggering pattern extraction, \toolname{} instructs an LLM to analyze the issue report and its corresponding fixing commit to extract well-formatted bug-triggering patterns.
In addition to raw issue and commit data, we also feed the LLM with the characteristics of the target bug, including its bug type, bug-triggering pattern category, and root cause category.
These labels are derived from the categorization established in our empirical study, providing the LLM with domain-specific prior knowledge about correctness bugs in \subject{}.
Given these inputs, we prompt the LLM to extract explicit, case-specific bug-triggering patterns aligned with each component of the standard PyTorch compiler test case (see Section~\ref{subsec:tool_test_format}).

Listing~\ref{lst:prompt_fault_triggering_pattern} outlines the structure and key components of our extraction prompt, including task objectives, input specifications, output formatting constraints, and extraction guidelines.

\input{listings/prompts/fault_triggering_pattern.tex}

\subsection{Mutation Rules Synthesis}\label{subsec:tool_mutation_synthesis}
We decompose the mutation rules for \subject{} test mutation into two fundamental components: a \textit{condition} and an \textit{action}.
The condition describes the scenarios under which a mutation is applied, whereas the action specifies the exact modification on test cases.
For illustration, taking a mutation rule that applies an in-place operation on tensor views as an example: its condition corresponds to checking whether the test case's model structure contains a view tensor, and its action corresponds to inserting an in-place operation on this view tensor (e.g., \texttt{view\_tensor[...] = 0}).
Following this structured formulation, the mutation rule synthesizer takes the extracted bug-triggering patterns as inputs and generates bug-triggering mutation rules for each component of the \subject{} test case.

Listing~\ref{lst:prompt_mutation_rule} presents the structure and key elements of our mutation rule synthesis prompt, including task description, input/output formats, and rule generation guidelines.
For each synthesized mutation rule, we also require the LLM to provide a concrete mutation example to facilitate the following test case mutation.
\input{listings/prompts/mutation_rule.tex}

\subsection{Test Case Mutation and Bug Detection}\label{subsec:tool_test_generation}
Lastly, \toolname{} employs an LLM as its test case mutator, which applies the synthesized mutation rules to seed test cases to generate new test cases.
Listing~\ref{lst:test_mutation_prompt} illustrates the structure of our test generation prompt.
This prompt instructs the LLM to apply the designated mutation rules to the seed test case, while enforcing predefined constraints to avoid syntactically or semantically invalid test cases and eliminate randomness that would lead to non-deterministic testing outcomes.

For bug detection, \toolname{} considers the standard computational assertions in the test case as its test oracle, including \texttt{torch.testing.assert\_close}.
Specifically, if any computational assertion in the test case fails, we flag the test case as triggering a potential correctness bug in \subject{}.

\input{listings/prompts/test_mutation.tex}

\subsection{Setup.}

\textbf{Configuration of LLMs.}
We utilize \texttt{gpt-4o} for pattern extraction and mutation rule synthesis, while adopting a smaller \texttt{gpt-4o-mini} model for test case mutation.
This choice is motivated by the relatively high computational cost associated with large-scale test case generation.
Consistent with the experimental settings in prior studies~\cite{jain2025livecodebench,can_emulating_semantic_translation,wu2025isolating,MR-Adopt}, we set the temperature parameter to 0.2.

\textbf{Seed Test Cases Collection.} 
Seed test cases are collected from the test suite of PyTorch version 2.8.0, the latest stable release at the time of our experimental implementation.
Specifically, we focused on \texttt{test/inductor/test\_torchinductor.py}, the primary test file for the PyTorch compiler, as determined through our analysis of the PyTorch test suite.
To facilitate subsequent mutation, we implement a custom test collector that automatically extracts structured components (including model structure, input data, compilation arguments, and testing pipelines) from raw test cases and converts them into a standardized format.
We then filter out unexecutable test cases caused by missing dependencies, incompatible system configurations, or other environment issues.
In total, 341 valid seed test cases are obtained from the original 475 test cases in the target test file.
Prior to the mutation process, we verify that none of these seed test cases induce correctness bugs in the latest version of PyTorch, ensuring that any newly detected bugs are indeed caused by our mutation process.

\textbf{Bug Detection and Validation.}
All test cases generated by \toolname{} are executed on the latest version of PyTorch to reduce the likelihood of detecting already known bugs.
We conduct further manual inspection on these test cases to filter out false positives, duplicated bugs, and previously reported bugs.
Finally, the newly identified correctness bugs are reported to the PyTorch development team for confirmation and fix.

\textbf{Environments.} 
\toolname{} was deployed on a server with a 32-core CPU and 500 GB of RAM\@.

\subsection{Summary of Detected New Correctness Bugs}
In total, \toolname{} detected 23 correctness bugs; all of these bugs have been triaged by developers and confirmed as previously unknown bugs, including 10 fixed in the latest version of PyTorch.
Moreover, \textbf{more than half of the detected correctness bugs (14/23) were even labeled as high-priority bugs by the development team}, which underscores the severity of these bugs and the practical value of \toolname{}.

\input{tables/Total-Bug.tex}

Table~\ref{tab:total_bugs} summarizes the distribution of the detected correctness bugs across different bug types, according to our empirical study.
Among the 23 correctness bugs detected by \toolname{}, memory-related bugs are the most prevalent (12 out of 23), followed by operator-related bugs (7 out of 23) and graph-related bugs (3 out of 23).
\toolname{} also detects one precision bug, which does not belong to the three main bug categories in our benchmark.

We next present concrete examples to illustrate the typical correctness bugs detected by \toolname{}, as well as the usefulness of our domain knowledge in guiding test case mutation for detecting these bugs.

\textbf{Memory-related Bugs.}
Among the 12 memory-related bugs detected by \toolname{}, most exhibit the same bug-triggering pattern, which is identified in our empirical study: \textit{in-place operations applied to tensor aliases}.
Listing~\ref{lst:example_detected_memory_error} presents a concrete example of a memory-related bug detected by \toolname{}, which has been marked as a high-priority bug by the development team~\cite{detected_memory_related}.
In this example, the DL model first constructs a view of the input tensors via concatenation, and subsequently performs in-place updates on this view.
These operations lead to incorrect outputs because the correct aliasing relationship is lost during the slicing and \texttt{index\_put} operations.
\toolname{} successfully identifies this bug by applying a synthesized mutation rule that applies in-place operations (e.g., indexed assignment) on reshaped tensors.
This rule guides its subsequent test mutation process to apply indexed assignment on a concatenated tensor, thus triggering this bug.

\input{listings/detected_error_examples/memory_related_error.tex}

\textbf{Operator-related Bugs.}
As for the 7 operator-related bugs detected by \toolname{}, these bugs cover multiple bug-triggering patterns identified in our empirical study.
Two are caused by single DL operators (including \texttt{logit} and \texttt{interpolate}) with \textit{non-default parameters or edge-case inputs}, while the remaining five are caused by specific operator sequences that trigger bugs during graph optimization.
Listing~\ref{lst:example_detected_operator_error} presents a concrete example of an operator-related bug identified by \toolname{}~\cite{detected_operator_related}.
In this example, the provided model triggers a constant folding optimization within \subject{}; however, this optimization fails to consider the corner case where the input is of \texttt{complex} data type, thereby resulting in incorrect model outputs.
This bug has been confirmed and fixed by the PyTorch development team after we reported it.

\input{listings/detected_error_examples/operator_related_error.tex}

\textbf{Graph-related Bugs.}
The three graph-related bugs detected by \toolname{} share a similar root cause: inconsistent random number generator (RNG) between the eager (non-compiled) and compiled modes of execution.
Specifically, Listing~\ref{lst:example_detected_graph_error} presents an example of such a graph-related bug identified by \toolname{}~\cite{detected_graph_related}.
In this instance, \subject{} incorrectly employs a different random number generator for the compiled model (which includes a \texttt{randn\_like} operation) when the input tensor is non-contiguous and has a size greater than 16.
A PR has been created by the development team to fix this bug, waiting to be merged to the latest version.

\input{listings/detected_error_examples/graph_related_error.tex}

%% file: listings/prompts/fault_triggering_pattern.tex
\begin{listing}
\begin{lstlisting}
|\underline{\textbf{Prompt}}|:
...
|\underline{Task}|:... summarize a case-specific bug-triggering pattern...
|\underline{Input}|: <issue report>, <fixing commit>, <bug type category>, <bug-triggering pattern category>, <root cause category>
|\underline{Output}|:
"Key Trigger Components": {
    "Model Structures": "", 
    "Input Features": "", 
    "Compilation Arguments": "", 
    "Testing Pipeline": ""
},
"Pattern Rationale": "...",
|\underline{Guidelines}|:
Identify the |\textbf{critical components that contribute to this bug}|
map them into the provided components ...
\end{lstlisting}
\caption{The structure of the bug-triggering pattern extraction prompt.}\label{lst:prompt_fault_triggering_pattern}
\end{listing}

%% file: listings/prompts/mutation_rule.tex
\begin{listing}
\begin{lstlisting}
|\underline{\textbf{Prompt}}|:
...
|\underline{Task}|: ... design targeted, template-aligned mutation plans that only focus on the mutable regions within this fixed seed template to trigger torch.compile correctness bugs...
|\underline{Input}|: <structured bug-triggering pattern>, <issue report>, <fixing commit>, <bug type category>, <root cause category>, <bug-triggering pattern category>
|\underline{Seed Template}|:
```
class Model(nn.Module):...
def GenInput():...
def testing_pipeline(Model, GenInput, compiler_args):...
compiler_args = {[torch compiler arguments declaration]}
...
```
|\underline{Output}|:
"Rule1": {
    "Target Mutable Component": "...",
    "Condition": "<The feature of seed code (e.g., 'if the model structure contains a view tensor')>",
    "Action": "<Concrete mutation action(s) to be applied on this seed code (e.g., 'Insert an in-place operation on this view tensor (e.g., view_tensor[...] = 0)')>",
    "Example Mutation": "<seed code snippet> -> <seed code snippet after mutation>",
},
"Rule2": {
    ...
},
...
```
|\underline{Guidelines}|: For each mutable component, if applicable, design a set of mutation rules (condition -> action pairs) to induce the bug-triggering pattern on diverse seed codes that follow the fixed template...
|\underline{Constraints}|: ...
|\underline{Example}|:...
\end{lstlisting}
\caption{The structure of the mutation rule synthesis prompt.}\label{lst:prompt_mutation_rule}
\end{listing}

%% file: listings/prompts/test_mutation.tex
\begin{listing}
\begin{lstlisting}
|\underline{\textbf{Prompt}}|:
...
|\underline{Input}|: <mutation plan>, <issue report>, <seed code>
|\underline{Output}|:
"Mutated Code": "<The mutated code after applying the mutation plan>",
"Explanation": "<Brief explanation of how the mutation plan was applied to the seed code>"
|\underline{Constraints}|:
-  Avoid breaking the constraints of the original seed code...
-  Avoid introducing syntax errors or semantic errors that are not related to the mutation plan...
-  Avoid introducing randomness that may lead to non-deterministic test results...
\end{lstlisting}
\caption{The structure of the test mutation prompt.}\label{lst:test_mutation_prompt}
\end{listing}

%% file: tables/Total-Bug.tex
\begin{table}[htbp]
\caption{New correctness bugs detected by \toolname{}}\label{tab:total_bugs}
\resizebox{0.7\linewidth}{!}{
\begin{tabular}{l|r|r|r|r@{}}
\toprule
Bug Type              & Reported & Confirmed & Fixed   & High-Priority   \\ \midrule
Memory-related Bugs & 12       & 12       & 7        & 8            \\ \midrule
Operator-related Bugs & 7        & 7        & 1        & 3            \\ \midrule
Graph-related Bugs & 3        & 3        & 1        & 2            \\ \midrule
Precision Bugs         & 1        & 1        & 1        & 1             \\ \midrule
Total                   & 23       & 23       & 10       & 14             \\ \bottomrule
\end{tabular}
}
\end{table}

%% file: listings/detected_error_examples/memory_related_error.tex
\begin{listing}
\begin{lstlisting}
|\textbf{\underline{Bug-Triggering Model}}|
def model(x, y):
    x = 2 * x
    c = torch.cat([x, y], dim=1)  -- create a view of `x' and `y' via concatenation
    c[:, [1, 0]] = c[:, [0, 1]]  -- |\textbf{\textcolor{red}{in-place update on the view, causing the bug!}}|
    return c[:, :2] + x
|\textbf{\underline{Model Input}}|
x1 = tensor([[0,1], [2,3]])
y1 = tensor([[0,1], [2,3]])
|\textbf{\underline{\textcolor{red}{Buggy Output}}}|
[[|\textbf{\textcolor{red}{4,0}}|],[|\textbf{\textcolor{red}{12,8}}|]]
|\textbf{\underline{\textcolor{codegreen}{Expected Output}}}|
[[|\textbf{\textcolor{codegreen}{2,2}}|],[|\textbf{\textcolor{codegreen}{10,10}}|]]
\end{lstlisting}
\caption{A simplified example of a \textit{Memory-related bug} detected by \toolname{}. In this example, \subject{} fails to correctly handle the in-place update on the view created by concatenation, leading to incorrect model outputs.}
\label{lst:example_detected_memory_error}
\end{listing}

%% file: listings/detected_error_examples/operator_related_error.tex
\begin{listing}
\begin{lstlisting}
|\textbf{\underline{Bug-Triggering Model}}|
def forward(self, x):
    y = torch.ones(2, 1, device=x.device, dtype=torch.complex64)
    c = x + y  -- this broadcast addition triggers constant folding optimization
    return c
|\textbf{\underline{Model Input}}|
x = tensor([[1.+0.j,1.+0.j],[1.+0.j,1.+0.j]])
|\textbf{\underline{\textcolor{red}{Buggy Output}}}|
tensor([[|\textbf{\textcolor{red}{2.+1.j,2.+1.j}}|],[|\textbf{\textcolor{red}{2.+1.j,2.+1.j}}|]])
|\textbf{\underline{\textcolor{codegreen}{Expected Output}}}|
tensor([[|\textbf{\textcolor{codegreen}{2.+0.j,2.+0.j}}|],[|\textbf{\textcolor{codegreen}{2.+0.j,2.+0.j}}|]])
\end{lstlisting}
\caption{A simplified example of a \textit{Operator-related bug} detected by \toolname{}. In this example, \subject{} fails to correctly conduct the broadcast addition of a complex tensor and a real tensor, leading to incorrect outputs.}
\label{lst:example_detected_operator_error}
\end{listing}

%% file: listings/detected_error_examples/graph_related_error.tex
\begin{listing}
\begin{lstlisting}
|\textbf{\underline{Bug-Triggering Model}}|
torch._inductor.config.fallback_random = True  -- Forces consistent random number generator between eager (non-compiled) and compiled model
def model(x):
    torch.manual_seed(0)
    return torch.randn_like(x)
|\textbf{\underline{Bug-Triggering Input}}|
x = torch.zeros((2,8)).permute(1,0)
|\textbf{\underline{\textcolor{red}{Buggy Behavior}}}|
Outputs between eager (non-compiled) and compiled model are inconsistent!
\end{lstlisting}
\caption{A simplified example of a \textit{Graph-related bug} Detected by \toolname{}. In this example, \subject{} fails to correctly capture the RNG state for the permuted tensor input, leading to inconsistent outputs between the eager (non-compiled) and compiled model.}
\label{lst:example_detected_graph_error}
\end{listing}

%% file: 7-Discussions.tex
\section{Discussions}\label{sec:discussions}
\subsection{Evaluation Scope for DL Compiler Testing Techniques}
When assessing the bug detection capabilities of the selected deep learning compiler testing techniques, our evaluation focuses specifically on their efficacy in identifying correctness bugs within \subject{}.
While these techniques may detect other bug types (e.g., crash bugs) in \subject{} during the evaluation process, these bugs are excluded from our analysis, as they lie outside the scope of this study.
Accordingly, our empirical findings on the performance of these testing techniques are specific to their correctness bug detection ability on \subject{}, and do not necessarily generalize to their performance in identifying other bug types in \subject{}.

\subsection{Threats to Validity}
Firstly, our study involves several manual steps, including manual filtering to identify real correctness bugs and manual categorization to analyze the characteristics of these bugs.
The subjective nature of these manual analyses may introduce bias into our results, thus posing a threat to the validity of our study.
To mitigate this threat, we adopt a systematic approach to ensure the reliability of our manual analysis.
For manual filtering, two authors independently analyze the collected issues with explicit filtering guidelines (see \S~\ref{subsec:error_collection_methodology}), and any inconsistent judgments are discussed until a consensus is reached.
For manual categorization, the first author initially categorizes the bug types, root causes, and bug-triggering patterns, after which the second author reviews all categorization results; inconsistencies are resolved to ensure agreement.

Secondly, our empirical study is grounded in an analysis of 116 correctness bugs related to \subject{}.
This introduces a potential threat to validity, as the generalizability of our findings beyond this curated dataset may be limited.
To mitigate this threat, we adopt a comprehensive data collection strategy that covers all issues with the keyword `torch.compile' in the past two years.
In addition, we further validate the practical value of our empirical evidence by constructing a proof-of-concept testing technique using our identified bug characteristics.
Notably, this technique has successfully detected 23 previously unknown correctness bugs in \subject{}, providing concrete evidence that our key findings extend beyond the originally collected 116 correctness bugs.

%% file: 8-Related-Works.tex
\section{Related Work}\label{sec:related-works}
\subsection{Related Work on DL Compiler Testing}
Existing testing approaches for DL compilers target generating effective test cases to trigger diverse code in DL compilers.
NeuRI~\cite{neuri} and NNSmith~\cite{nnsmith} are two grammar-based generation fuzzers, targeted at generating diverse yet valid computational graphs for testing.
These techniques use predefined (or inductively learned) operator constraints to generate valid computational graphs, then apply dedicated mutation strategies to produce diverse graphs.
Opera~\cite{opera} targets bugs in the model loading stage by migrating operator-level test cases from DL frameworks (e.g., PyTorch API test cases) to DL compilers such as TVM\@.
With the rise of LLMs, recent work has explored using these models for DL compiler testing.
Among them, WhiteFox~\cite{whitefox} adopts an LLM-based static analysis technique to analyze the optimization logics in DL compilers, leveraging these insights to generate optimization-aware test cases.
DeepConstr~\cite{deepconstr} is a constraint-guided fuzzer that utilizes LLMs to generate more complete constraints, thereby extending the search space of valid test cases.
All these techniques have made significant progress in test diversity.
However, they are not equipped with domain knowledge of correctness bugs in \subject{}, and thus may not be effective at generating correctness bug-triggering test cases as demonstrated in our study in Section~\ref{sec:benchmark}.
In contrast, our study focuses on characterizing the correctness bugs in \subject{} and derives actionable insights for effective bug detection, which can complement existing testing techniques and guide the design of more effective testing techniques for \subject{}.

Beyond DL compiler testing, several empirical studies have been dedicated to characterizing bugs in DL compilers.
Shen et al.~\cite{dlcompilerbugstudy1} analyzed the root causes, symptoms, and faulty stages of bugs in three DL compilers, and further proposed practical guidelines for debugging and testing DL compilers.
Du et al.~\cite{dlcompilerbugstudy2} categorized DL compiler bugs into five root-cause types, and systematically investigated their distribution, impacts, and fixing times.
Wu et al.~\cite{empirical_study_on_TVM} focused on bug reports in TVM, exploring the core challenges in TVM usage and development, and further analyzed the impacts and symptoms of the reported bugs.
Compared with existing empirical studies, our work is the first to systematically investigate correctness bugs in \subject{}, which represent a critical yet understudied problem in \subject{}.

\subsection{Related Work on Correctness Bugs in AI Infrastructure}
Driven by the severe impact of bugs in AI systems, correctness bugs within AI infrastructure have received growing research attention.
A series of studies have been proposed to characterize and detect such correctness bugs across the AI ecosystem.
TrainCheck~\cite{jiang2025trainingconfidencecatchingsilent} automatically infers specialized invariants for DL training and uses these invariants to proactively detect silent training errors during model execution.
Tambon et al.~\cite{silenterrorinDLframework} report an empirical study on silent bugs in Keras and TensorFlow, revealing their user-level impacts, root causes, and associated faulty components.
Hong et al.~\cite{silentbugs_in_pytorch} further investigate the silent bugs in user-developed PyTorch programs, analyzing their symptoms, underlying causes, and common bug patterns.
However, existing research on correctness bugs in AI infrastructure mainly focuses on user-level code or DL library implementations.
In contrast, our work targets DL compilers, a critical but understudied component in the AI stack with distinct architectural design and bug characteristics compared to user-level code and DL libraries.

%% file: 9-Conclusion.tex
\section{Conclusion}
In this paper, we present the first systematic study on the correctness bugs in the PyTorch compiler, \subject{}.
Through an in-depth analysis of 116 real-world correctness bugs, we categorize common bug types, root causes and triggering patterns of these bugs.
We further assessed the performance of five existing DL compiler testing techniques in detecting these bugs, understanding their limitations and providing insights for improving testing techniques for this task.
At last, we propose a proof-of-concept testing technique, \toolname{}, that leverages the bug characteristics identified in our empirical study to detect correctness bugs in \subject{}.
Our evaluation shows that \toolname{} effectively detects 23 new correctness bugs in \subject{}, all these bugs have been confirmed or fixed by the PyTorch development team, including 14 labeled as high-priority bugs.
The effectiveness of \toolname{} in detecting new correctness bugs in \subject{} demonstrates the practical value of our findings and the potential of our approach for improving the reliability of DL compilers.